\documentstyle[emulateapj,apjfonts,epsfig,graphicx]{article} 

\def\bri{BRI\thinspace0021$-$0214}
\def\bris{BRI\thinspace0021}

\def\pc{PC\thinspace0025+0447}
\def\pcs{PC\thinspace0025}

\def\lp{LP\thinspace944-20}

\def\denis{DENIS-P J1228.2$-$1547}
\def\deniss{DENIS\thinspace1228}

\def\mass{2MASP J0345432+254023}
\def\masss{2MASP\thinspace0345+25}

\def\sdss{SDSS J134646.45-003150.4}
\def\sdsss{SDSS\thinspace1346}

\def\tvlm{TVLM\thinspace513-46546}
\def\tvlms{TVLM\thinspace513}

\def\gd{GD\thinspace165B}
\def\crbr{CRBR\thinspace15}

\begin{document}
 
\title{\large Flaring Up All Over --- Radio Activity in
Rapidly-Rotating Late M and L Dwarfs}

\author{E. Berger} 
\affil{Division of Physics, Mathematics, and Astronomy, 
California Institute of Technology 105-24, Pasadena, CA 91125\\
ejb@astro.caltech.edu}

\begin{abstract}
\noindent We present Very Large Array observations of twelve
late M and L dwarfs in the Solar neighborhood.  The 
observed sources were chosen to cover a wide range of physical
characteristics --- spectral type, rotation, age, binarity, and X-ray 
and H$\alpha$ activity --- to determine the role of these properties
in the production of radio emission, and hence magnetic fields.
Three of the twelve sources, \tvlm{}, 2MASS J0036159+182110, and
\bri{}, were observed to flare and also exhibit persistent emission,
indicating that magnetic activity is not quenched at the bottom of the
main sequence.  The radio emission extends to spectral type L3.5, and
there is no apparent decrease in the ratio of flaring
luminosities to bolometric luminosities between M8$-$L3.5.  Moreover,
contrary to the significant drop in persistent H$\alpha$ activity
beyond spectral type M7, the persistent radio activity appears to
steadily increase between M3$-$L3.5.  Similarly, the radio emission
from \bri{} violates the phenomenological relations between the radio
and X-ray luminosities of coronally active stars, hinting that radio
and X-ray activity are also uncorrelated at the bottom of the main
sequence; an even stronger violation was found for the brown dwarf
LP944-20.  The radio active sources that have measured rotational
velocities are rapid rotators, $v{\rm sin}i>30$ km
sec$^{-1}$, while the upper limits on radio activity in
slowly-rotating late M dwarfs ($v{\rm sin}i<10$ km sec$^{-1}$) from
this survey and from the literature are lower than these detections.
These observations provide tantalizing evidence that rapidly-rotating 
late M and L dwarfs are more likely to be radio active.  
This possible correlation is puzzling given that the observed 
radio emission requires sustained magnetic fields of $\sim 10-10^3$ G and
densities of $\sim 10^{12}$ cm$^{-3}$, indicating that the active
sources should have slowed down considerably due to magnetic braking. 
\end{abstract}


\section{Introduction}\label{sec:intro}

Activity in dwarf stars, as measured through the ratios of H$\alpha$
and X-ray luminosities to bolometric luminosities, drops significantly
in spectral types later than M7, and is possibly dominated by flares
(Liebert et al. 1999; Reid et al. 1999; Gizis et al. 2000; Kirkpatrick
et al. 2000)\nocite{lkr+99,rkg+99,gmr+00,krl+00}.  In addition, the
dependence of activity on rotation appears to break down in sources
with spectral type later than early-mid M, so that even rapidly
rotating sources ($v{\rm sin}i\gtrsim 20$ km sec$^{-1}$) show a 
decrease in emission (Basri \& Marcy 1995; Tinney \& Reid 1998)
\nocite{bm95,tr98}.  On the other hand, the transition from a
radiative-convective ($\alpha\Omega$) to a turbulent ($\alpha^2$)
dynamo (Durney, De Young \& Roxburgh 1993)\nocite{ddr93} around the
convective mass limit of 0.3 M$_\odot$ (spectral type $\sim {\rm M}3$)
is not accompanied by a drop in activity (Hawley, Gizis \& Reid
1996)\nocite{hgr96}, indicating that in early M dwarfs the
turbulent dynamo can sustain active coronae.  It has been suggested
that the observed shift to weak flaring activity in late M and L
dwarfs is possibly due to a change in the nature of the dynamo, or
decreased photospheric ionization levels.  If radio activity follows
the same pattern, then late M and L dwarfs would exhibit at most
only weak flaring, and possibly weaker quiescent radio emission.     

In addition, Guedel \& Benz (1993)\nocite{gb93}, and Benz \&
Guedel (1994)\nocite{bg94} have shown that for a broad range of
coronally active stars (up to spectral type M7) there  is a
correlation between the radio and X-ray luminosities.  Simply, the
quiescent luminosities are related by $L_{\rm rad,q}\approx L_{\rm
X,q}/10^{15.5}$ Hz$^{-1}$, while the relation for flaring luminosities
is slightly non-linear with $L_{\rm rad,f}\approx L_{\rm X,f}/10^{15.5}$
Hz$^{-1}$ for radio flares with $L_{\rm rad}\sim 10^{14}$ erg sec$^{-1}$
Hz$^{-1}$.  These relations are explained in terms of a heating
process in which there is a causal relation between the radio-emitting
electrons and the X-ray emitting thermal plasma.  The extension of the
Guedel--Benz relations to spectral types later than M7
based on X-ray observations (Neuh{\"a}user et al. 1999; Rutledge et
al. 2000) predicts radio emission well below the detection limit of
the Very Large Array.  Past observations seemed to confirm
this prediction (Krishnamurthi, Leto \& Linsky 1999)\nocite{kal+99}.

The recent detections of an X-ray flare (Rutledge et al. 2000) and
persistent and flaring radio emission (Berger et al. 2001; hereafter
B01\nocite{bbb+01}) from the brown dwarf \lp{} (spectral type M9) pose
a serious 
challenge to these ideas, since they demonstrate that late dwarfs can
exhibit strong activity closely related to magnetic fields.  The
quiescent radio emission in particular indicates that the magnetic
activity is persistent, contrary to H$\alpha$ and X-ray
observations of this source (Rutledge et al. 2000).  Moreover, the
observed emission violates both Guedel--Benz relations by
approximately four orders of magnitude despite the fact that \lp{} is
only an M9 source, hinting at a possible breakdown in the relation
between radio and X-ray activity between spectral types M7 and M9.
Finally, the rapid rotation of \lp{}, $v{\rm sin}i\approx 30$ km
sec$^{-1}$, is puzzling given that the radio emission requires the
presence of magnetic fields and ionized material, which should lead to
magnetic braking. 

These observations raise several crucial questions: Is \lp{} a unique
object, or do other late-type stars and brown dwarfs show similar
levels of radio activity?  What is the underlying mechanism that gives
rise to the emission, and is it similar to the mechanism in early M
dwarfs?  Is the same process responsible for the observed H$\alpha$
and X-ray flares, and in some cases persistent emission?  Is the
activity correlated with particular physical properties? 

To address these questions, we have undertaken a survey of twelve
late-type dwarfs (ranging from M5 through T6) that span a wide range of 
properties.  The
purpose of this survey is to search for flaring and/or persistent
radio emission, and to investigate whether this emission, and
presumably magnetic field production, are correlated with particular
physical properties of the sources.   

We summarize the properties of the survey sources in \S\ref{sec:ss}. 
The radio observations and data reduction are detailed in
\S\ref{sec:obs}.  In \S\ref{sec:res} we show that three of the twelve
sources exhibit radio emission.  We analyze the radio emission, show that 
at least in one case it violates the Guedel--Benz relations, and estimate 
the magnetic field
strengths and coronal densities in \S\ref{sec:flare}.  We also discuss the 
implications for coronal heating mechanisms (\S\ref{sec:heat}).  Finally, in
\S\ref{sec:disc} we compare the results of the survey to previous radio surveys  
of M and L dwarfs, as well as LP944-20, and we draw preliminary conclusions about 
possible correlations between radio activity and physical parameters.

\section{Target Selection}
\label{sec:ss}

We observed twelve late M and L dwarfs, two of which are
confirmed brown dwarfs, ranging in distance from $\approx 8-160$ pc,
and spanning a wide range of ages, rotation, activity, and spectral
types.  Below and in Table~\ref{tab:sources} we summarize the main
properties of each source in the survey.

\noindent{\bf \crbr{}}: This source is a brown dwarf candidate in the 
$\rho$ Oph molecular cloud (Wilking, Greene \& Meyer
1999)\nocite{wgm99}. The mass and age of \crbr{},  $M\lesssim 0.1$
M$_\odot$ and $t\lesssim 3$ Myr, are estimated by a comparison to
evolutionary models and isochrones (Burrows et al. 1997; Wilking,
Greene \& Meyer 1999)\nocite{bmh+97,wgm99}.  Wilking et al. (1999) 
claim that while the age of this source is secure relative to other
sources in the $\rho$ Oph cloud, the absolute scale is
model-dependent.  Still, this source is the youngest in our sample.
The temperature and bolometric luminosity of \crbr{} are estimated at
$2900-3000$ K and ${\rm log}(L_{\rm bol}/L_\odot)\approx -1.0$ to
$-1.3$  (Luhman \& Rieke 1999; Wilking, Greene \& Meyer
1999)\nocite{lr99,wgm99}.  In a recent  observation  of the $\rho$ Oph
molecular cloud with Chandra (Imanishi, Tsujimoto \& Koyama
2001)\nocite{itk01} an upper limit of $L_{\rm X}<1.1\times 10^{28}$
erg sec$^{-1}$ was found for this source, with ${\rm log}(L_{\rm
X}/L_{\rm bol})<-4.2$, well below  the X-ray saturation limit, ${\rm
log}(L_{\rm X}/L_{\rm bol})\approx -3$.  In addition, \crbr{} exhibits
excess emission at 2.2 $\mu$m (Wilking, Greene \& Meyer
1999)\nocite{wgm99}, and has a flux density of 40 mJy at 1.3 mm
(Motte, Andre \& Neri 1998)\nocite{man98}.  These observations
indicate that it is associated with a circumstellar envelope or an
accretion disk. Thus, radio emission from this source can be due to
accretion or deuterium burning. 

\noindent{\bf LHS\,2243}: This source is a slowly rotating, $v{\rm
sin}i<5$ km sec$^{-1}$, M8 dwarf.  It shows evidence for weak
persistent H$\alpha$ emission, as well as H$\alpha$ flares, with an
increase of $\sim 30$ in the line equivalent width (Mart{\' i}n,
Rebolo \& Magazzu 1994; Gizis et al. 2000)\nocite{mrm94,gmr+00}.  The
temperature of LHS\,2243 is estimated to be $\sim 2900$ K (Mart{\'
i}n, Rebolo \& Magazzu 1994) based on the calibration of Kirkpatrick
et al. (1993)\nocite{kmh+93}.  This calibration provides the highest 
temperatures among the published temperature conversions (Mart{\' i}n,
Rebolo \& Magazzu 1994).  Bessell (1991)\nocite{bes91} suggested that
this source is possibly as young as a ${\rm few}\times 10^8$ yr, based on
its smaller proper motion compared to bluer stars.  For an age of
$\sim 10^9$ yr, the inferred mass, based on an upper limit on Li
abundance and several evolutionary models is $\sim 0.06-0.07$
M$_\odot$ (Mart{\' i}n, Rebolo \& Magazzu 1994).  If the source is
younger than $10^9$ yr, the mass would be slightly higher.
  
\noindent{\bf \tvlm{}}:  This is a nearby M8.5 dwarf (Tinney 1993;
Tinney et al. 1995; Reid et al. 2001)\nocite{tin93,trg+95,rbc+01}.
\tvlm{} (hereafter \tvlms{}) is the fastest rotator in our sample with
$v{\rm sin}i\approx 60$ km sec$^{-1}$, but it has only weak H$\alpha$
emission with equivalent width of 2.5\AA{} (Mart{\' i}n, Rebolo \&
Magazzu 1994).  Based on $1-2.5$ $\mu$m spectroscopy, and synthetic
spectra fitting, Leggett et al. (2001)\nocite{lag+01} find ${\rm
log}(L_{\rm bol}/L_\odot)\approx -3.65$ and $T\approx 2200$ K for this
source.  We note however, that their temperature scale provides
consistently lower temperatures relative to other estimates (see notes
on LHS\,2065 below).

\noindent{\bf LHS\,2065}: This source has spectral type M9 and
relatively slow rotational velocity, $v{\rm sin}i\approx 9$ km
sec$^{-1}$ (Tinney \& Reid 1998; Mart{\' i}n et al. 1999; Reid et
al. 2001)\nocite{tr98,mdb+99,rbc+01}.  It is active in H$\alpha$,
exhibiting strong flares with a recurrence of $< 0.03$ hr$^{-1}$,
and possibly more frequent weak flares at a rate $\sim 0.5$ hr$^{-1}$ 
(Mart{\' i}n \& Ardila 2001)\nocite{ma01}.  There are several
estimates of the effective temperature of this source in the
literature, ranging from $2100-2600$ K.  Leggett et
al. (2001)\nocite{lag+01} provide an estimate of 2100 K (see notes  on
\tvlms{} above).  Another estimate, based on near infrared photometry,
provides a temperature of 2300 K (Leggett, Allard \& Hauschildt 
1998; Mart{\' i}n et al. 1999)\nocite{lah98,mdb+99}, while Mart{\'i}n,
Rebolo \& Magazzu (1994) provide  a value of 2630 K (see notes on
LHS\,2243 above).  The luminosity of LHS\,2065 is  estimated to be
${\rm log}(L/L_{\rm bol})\approx -3.5$ (Leggett et al. 2001), and its 
mass is estimated at $\sim 0.06-0.1$
M$_\odot$  (Mart{\' i}n, Rebolo \& Magazzu 1994)\nocite{mrm94} 
which places it close to the brown dwarf cutoff.

\noindent{\bf \bri}: This nearby, M9.5 dwarf is a rapid rotator,
$v{\rm sin}i\approx 40$ km sec$^{-1}$ (Basri \& Marcy 1995; Tinney \&
Reid 1998)\nocite{bm95,tr98}, but it exhibits no persistent H$\alpha$
emission (Tinney \& Reid 1998)\nocite{tr98}.  Reid et
al. (1999)\nocite{rkg+99} discovered an H$\alpha$ flare from this
source, with a peak emission three times lower than the mean quiescent
emission of early M dwarfs.  In addition, \bri{} (hereafter
\bris{}) has an upper limit on X-ray emission of ${\rm log}(L_X/
L_{\rm bol})<-4.7$, well below the X-ray saturation limit of ${\rm
log}(L_X/L_{\rm bol})\approx -3$ for X-ray active stars (Neuh{\"a}user
et al. 1999)\nocite{nbc+99}.  Leggett et al. (2001) estimate the
temperature of this source at 2100 K, and its luminosity at ${\rm 
log}(L_{\rm bol}/L_\odot)\approx -3.4$.  Finally, Mart{\'i}n, Basri \&
Zapatero Osorio (1999)\nocite{mbz99} find an upper limit on Li EW of
$<0.08$ \AA.  The rapid rotation, flaring emission, and lack  of
persistent H$\alpha$ and X-ray emission indicate that this source is
an analog of \lp{}.    

\noindent{\bf \pc{}}: This source is possibly a young ($<1$ Gyr) brown  
dwarf (M$<0.06$ M$_\odot$; Mart{\' i}n, Basri, \& Zapatero Osorio 
1999)\nocite{mbz99}, based on the detection of Li with EW$=1.0\pm 0.3$
\AA.  \pc{} (hereafter \pcs{}) exhibits very strong and persistent
H$\alpha$ emission (Graham et al. 1992; Mould et
al. 1994)\nocite{gmg+92,mco+94}.  Burgasser et
al. (2000)\nocite{bkr+00}, on the other hand, suggest that \pcs{} is
an interacting binary in order to explain the Li and H$\alpha$
observations. The rotational velocity of  \pcs{} is surprisingly low,
$v{\rm sin}i\approx 13\pm 3$ km sec$^{-1}$, given the strong level of
H$\alpha$ activity.  In addition, \pcs{} has an upper limit on X-ray
emission of ${\rm log}L_{\rm X}<27.2$ erg sec$^{-1}$ 
(Neuh{\"a}user et al. 1999)\nocite{nbc+99}.  The bolometric luminosity
of this source is estimated at ${\rm log}(L_{\rm bol}/L_\odot)\approx
-3.7$ (Schneider et al. 1991)\nocite{sgs+91}, giving ${\rm log}(L_X/
L_{\rm bol})<-2.7$.  The slow rotation and strong H$\alpha$ activity
place \pcs{} on the opposite end of the activity-rotation spectrum
from \bris{} and \lp. 

\noindent{\bf \mass{}}: This source is an isolated L0 dwarf, but it is
still not clear whether it is a low mass star or a brown dwarf younger
than $\sim 1$ Gyr (Kirkpatrick, Beichman \& Skrutskie
1997)\nocite{kbs97}.  Leggett et al. (2001) estimate the luminosity of
\mass{} (hereafter \masss{}) at ${\rm log}(L_{\rm bol}/L_\odot)\approx
-3.6$, with an effective temperature of 2000 K (see notes on \tvlms{}
above).  \masss{} has an upper limit on H$\alpha$ emission of ${\rm
log}(L_{\rm H\alpha}/L_{\rm bol})\lesssim -5.6$. 

\noindent{\bf 2MASSW J0036159+182110}: This source is a 
recently-discovered L3.5 candidate brown dwarf (Kirkpatrick et al. 
2000; Gizis et al. 2000; Reid et al. 2000)\nocite{krl+00,gmr+00,rkg+00}.  
2MASSW J0036159+182110 (hereafter 2MASS\,0036+18) shows no evidence of Li
or H$\alpha$, with upper limits of EW$<0.1$\AA{} (Reid et al. 2000;
Kirkpatrick et al. 2000)\nocite{rkg+00,krl+00}.  The temperature of 
2MASS\,0036+18 is estimated to be 1800 K based on the spectra and
models of Leggett et al. (2001).

\noindent{\bf \gd{}}: This source is a rapidly-rotating, $v{\rm
sin}i\approx 37$ km sec$^{-1}$, L4 dwarf orbiting a
white dwarf primary (Becklin \& Zuckerman 1988)\nocite{bz88}.  
Kirkpatrick et al. (1999)\nocite{kab+99} used model atmospheres,
including the effects of condensation and dust opacities, to estimate
the temperature of this source, $T\approx 1900$ K, and in addition
used the cooling age of the white dwarf primary to estimate the age of
\gd{} at $1.2-5.5$ Gyr.  Based on these two quantities they conclude
that this source is likely a brown dwarf.  The luminosity of \gd{} is
estimated at ${\rm log}(L_{\rm bol}/L_\odot)\approx -4.1$ (Leggett et
al. 2001).  Along with \denis{} (see below), and possibly \pcs{}, 
observations of \gd{} can be used to test the effect of binarity on 
radio emission from late dwarfs, and the 4'' separation between the 
binary members allows us to pinpoint the source of any radio emission 
from the system.   

\noindent{\bf 2MASSW J1507476$-$162738}: This source is a nearby L5
dwarf (Reid et al. 2000,2001)\nocite{rkg+00,rbc+01}.  2MASSW
J1507476$-$162738 (hereafter 2MASS\,1507$-$16) has only upper limits
on Li and H$\alpha$ of $<0.1$ \AA{} and $<0.5$ \AA, respectively
(Kirkpatrick et al. 2000; Reid et al. 2000)\nocite{rkg+00,krl+00}.  

\noindent{\bf \denis{}}: This source is an L5 brown dwarf
exhibiting relatively slow rotation, $v{\rm sin}i\approx 11$ km
sec$^{-1}$, and no detectable H$\alpha$ emission, ${\rm
log}(L_{H\alpha}/L_{\rm bol})\lesssim -5.3$ (Mart{\' i}n et
al. 1997; Delfosse et al. 1997; Tinney, Delfosse \& Forveille
1997)\nocite{mbd+97,dtf+97,tdf97}.  The bolometric luminosity of
\denis{} (hereafter \deniss{}) is estimated to be ${\rm log}(L_{\rm
bol}/L_\odot)\approx -4$ to $-4.3$ (Leggett et al. 2001; Mart{\'i}n et
al. 1997; Delfosse et al. 1997; Tinney, Delfosse \& Forveille 1997),
and its effective temperature is 1800 K (Leggett et al. 2001).  In
addition, \deniss{} has an upper limit on X-ray emission of ${\rm
log}L_{\rm X}< 27.3$ erg sec$^{-1}$, and ${\rm log}(L_X/
L_{\rm bol})<-2.3$.  Based on the detection of
Li and the estimated effective temperature, the age and mass of 
\deniss{} are estimated to be M$<0.065$ M$_\odot$ and $t<1$ Gyr
(Tinney, Delfosse \& Forveille 1997).  Finally, Mart{\' i}n, Brandner
\& Basri (1999)\nocite{mbb99} have shown that \deniss{} is in fact a
brown dwarf binary with a projected separation of approximately 5 AU.
While our observations cannot resolve the system, we can study the
combined emission, and effect of binarity on the production of this
emission.   

\noindent{\bf \sdss{}}: This is the only methane (spectral class T6;
Burgasser et al. 2001; Geballe et al. 2001)\nocite{bkb+01,gkl+01}
brown dwarf in the sample.  It has a slightly higher effective
temperature than that of two other known methane brown dwarfs,
GL\thinspace229B and SDSS J162414.37+002915.6 (Tsvetanov et
al. 2000)\nocite{tgz+00}, and it is inactive in H$\alpha$, with 
${\rm log}(L_{\rm H\alpha}/L_{\rm bol})\lesssim -5.3$ (Burgasser et
al. 2000)\nocite{bkr+00}.  The bolometric luminosity of \sdss{}
(hereafter \sdsss{}) is ${\rm log}(L_{\rm bol}/L_\odot)\approx -5.3$
(Burgasser et al. 2001; Tsvetanov et al. 2000).  At a distance of only
$\sim 11$ pc (Tsvetanov et al. 2000)\nocite{tgz+00} \sdsss{} is an
excellent candidate for the detection of radio emission from this
class of objects.

\section{Radio Observations}
\label{sec:obs}

Very Large Array (VLA\footnotemark\footnotetext{The VLA is operated by
the National Radio Astronomy Observatory, a facility of the
National Science Foundation operated under cooperative agreement by
Associated Universities, Inc.}) observations were conducted at 8.46
GHz in the standard continuum mode with $2\times 50$ MHz contiguous
bands.  A log of all observations is given in Table~\ref{tab:vla}.
Following the detection of emission from \bris{} at 8.46 GHz, we also
observed this source at 4.86 GHz.  In all observations we used the
extra-galactic sources 3C\thinspace48 (J0137+331) and 3C\thinspace286
(J1331+305) for flux calibration, while the phase was monitored using
calibrator sources within $\sim 5^\circ$ of the survey sources.   

The first set of observations was undertaken in the A and
A$\rightarrow$B configurations.  Unfortunately, half the data in these
observations had to be discarded due to a malfunction of the lower
Intermediate Frequency channel, and the resulting rms noise levels are
therefore higher than the theoretical values by $\approx \sqrt{2}$.
We repeated a subset of these observations over a period of
approximately three weeks in the CnB configuration.  The final set of
observations, with five new targets was undertaken in the DnC
configuration. 

The data were reduced and analyzed using the Astronomical Image
Processing System (AIPS, release 31DEC1999; Fomalont
1981)\nocite{fom81}.  Both the visibility data and maps were inspected
for data quality and noisy points were removed.  To search for
flares, we constructed lightcurves using the following method.  We
removed all the bright background sources in each image using the
AIPS/IMAGR routine to CLEAN the region around each source, and the
AIPS/UVSUB routine to subtract the resulting source models from the
visibility data.  We then plotted the real part of the complex
visibilities at the position of each target source as a function of
time using the AIPS/UVPLT routine.  The background source subtraction
is necessary since the side-lobes of these sources, and the change in
the synthesized beam shape during the observation result in flux
variations over the map, which may contaminate any real variability or
generate false variability.

\begin{figure} 
\plotone{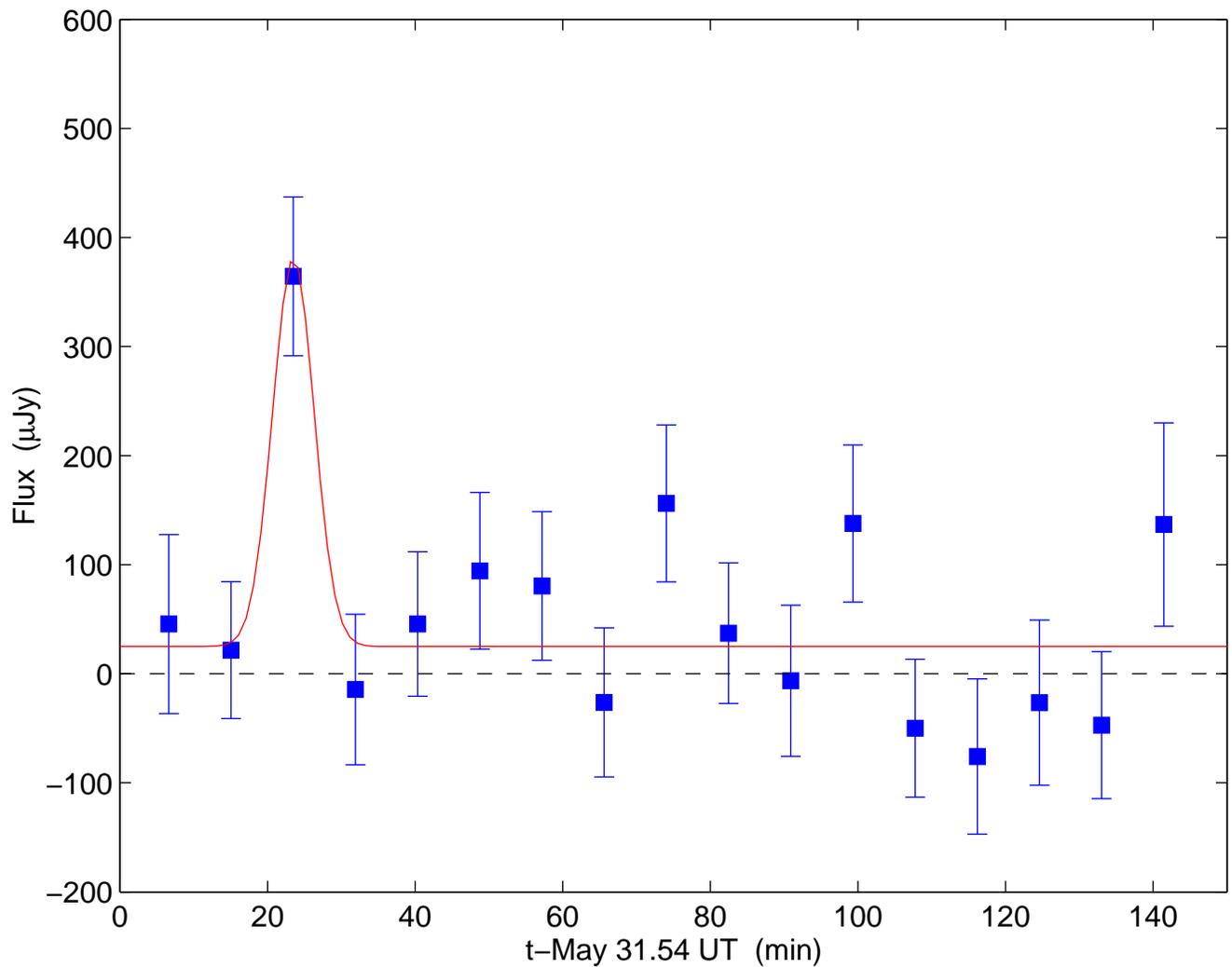}
\caption[]{Lightcurve of the 8.46 GHz emission from \bris{} on 2001,
May 31.54 UT, with 17 8-min bins.  We find a single flare with a
statistical significance of $4.5\sigma$.  The solid line is a Gaussian
model (Equation~\ref{eqn:model}).  The fraction of circular
polarization near the peak is $30\pm 20\%$ (see Table~\ref{tab:prop}).   
\label{fig:bri}}
\end{figure}

The uncertainties in the resulting lightcurves
(Figures~\ref{fig:tvlm}--\ref{fig:bri}) are dominated by system
thermal noise.  The system temperature includes contribution from
background sources, the sky, and hardware (e.g. receivers).  At the
observing frequency used in this survey, 8.46 GHz, the dominating term
is the receiver noise.  As a result, there are no significant
systematic effects since in all cases the sources were located within
a few arcseconds of the array pointing center, and the background
source subtraction successfully removed possible contaminating
sources.  In fact, the plotted uncertainties, as well as the rms noise 
from the maps, are within a few percent of the theoretical noise
estimates for the VLA at 8.46 GHz. 

To further test the veracity of the resulting lightcurves and
uncertainty estimates we constructed equivalent lightcurves for random
positions on each map.  We find that on average the fluctuations in
these lightcurves do not exceed $2\sigma$, and the noise estimates 
are similar.

\section{Results}
\label{sec:res}

\begin{figure} 
\plotone{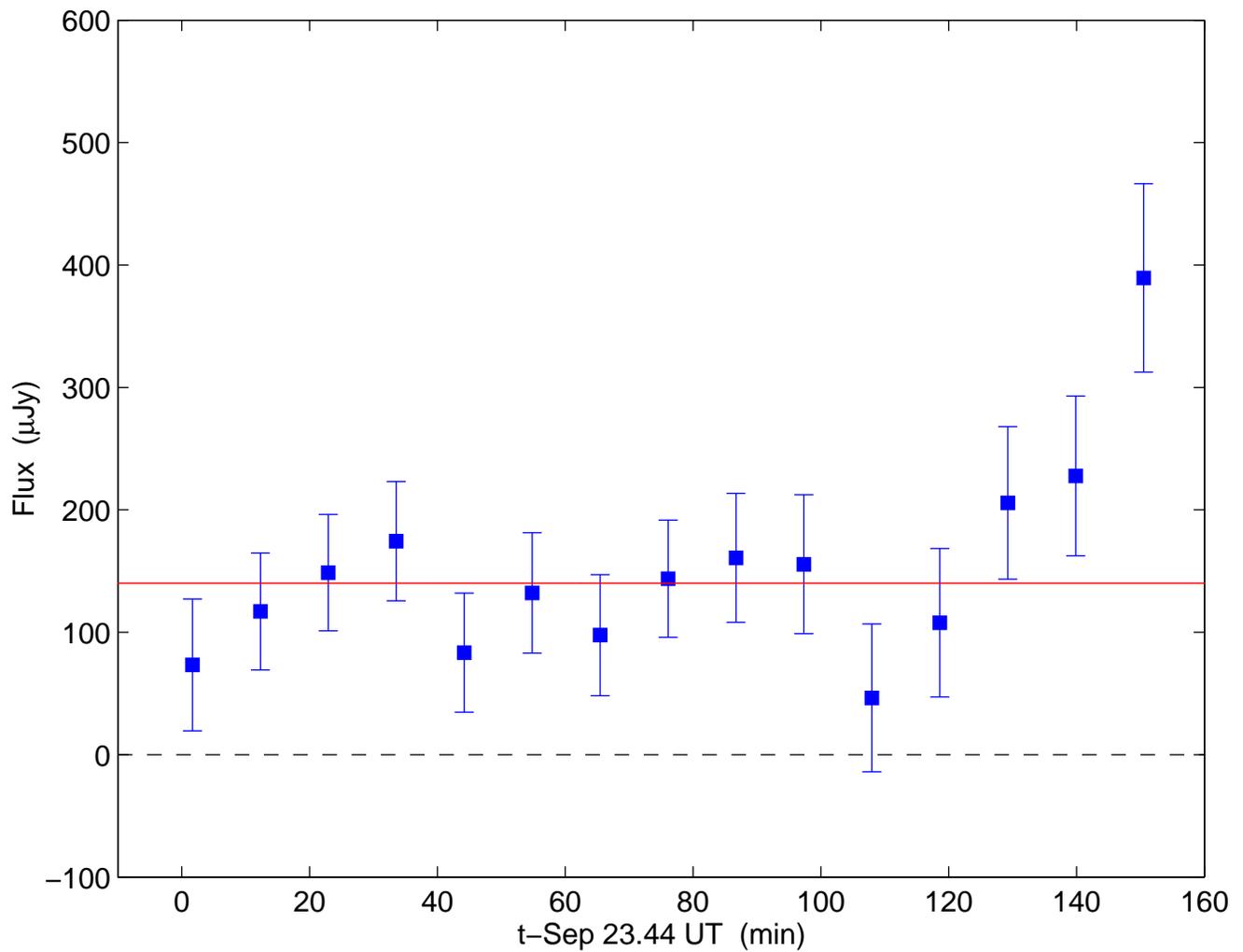}
\caption[]{Lightcurve of the 8.46 GHz emission from 2MASS\,0036+18 on
2001, Sep 23.44 UT, with 15 8.5-min bins.  We find only persistent
emission, with a marginal indication for re-brightening over the last
thirty minutes of the observation.  This possibly indicates the rise
of a strong flare.  The solid line is a fit to a constant source.
\label{fig:mass1}}
\end{figure}

We detected flares and persistent emission from \tvlms{},
2MASS\,0036+18, and, with lower significance, from \bris{}.
Figures~\ref{fig:tvlm}--\ref{fig:bri} show the lightcurves for
observations of these sources.  The peak flux densities range from
$\approx 360$ $\mu$Jy in \bris{} to $\approx 980$ $\mu$Jy in \tvlms{},
and the flare timescales (FWHM) range from $\approx 6$ min in \bris{}
to $\approx 20$ min in 2MASS\,0036+18.  The resulting duty cycles,
defined as the ratio of the time the sources produce flares to the
total observing time, are $\approx 2-10\%$.  In the case of \bris{},
the value of 2\% is consistent with a duty cycle of $\lesssim 7\%$ for
H$\alpha$ flares (Reid et al. 1999)\nocite{rkg+99}.  The duty cycles
of all three sources are consistent with an average H$\alpha$ flare
duty cycle of $\sim 7\%$ for late M dwarfs (Gizis et
al. 2000)\nocite{gmr+00}.  

The fraction of circular polarization
during the flares is significant in \tvlms{}, $f_{\rm circ}\approx
66\pm 4\%$, and 2MASS\,0036+18, $f_{\rm circ}\approx -62\pm 5\%$, but
is somewhat lower in \bris{}, $f_{\rm circ}\approx 30\pm 20\%$ (see
Figures~\ref{fig:tvlm} and \ref{fig:mass2}, and
Table~\ref{tab:prop}). The negative value for 2MASS\,0036+18 indicates
that the emission is left-handed circularly polarized. 

In addition, we clearly detect persistent emission from \tvlms{} and
2MASS\,0036+18, with flux densities of $\approx 190$ $\mu$Jy and
$\approx 240$ $\mu$Jy, respectively.  In the initial observation of
2MASS\,0036+18 the persistent flux density was only $\approx 130$ $\mu$Jy.
The detection of persistent emission from \bris{} is marginal, with an
average value of $40\pm 13$ $\mu$Jy at 8.46 GHz from the combined
observations is January and May.  At 4.86 GHz the flux from this
source is $40\pm 18$ $\mu$Jy, and it is not clear whether persistent
emission is detected.

In addition to the change in persistent flux between the two
observations of\\ 2MASS\,0036+18, the persistent emission from this
source and \tvlms{} appears to vary within each observation.
In the case of 2MASS\,0036+18 the source brightened over the last
60 min of observation on Oct 9, possibly indicating a second strong
flare (Figure~\ref{fig:mass2}).  The sharp decrease in flux over the
first 30 min of the observation could be the exponential tail of
an earlier strong flare.  An increase in flux was also observed towards
the end of the observation from Sep 23 (Figure~\ref{fig:mass1}).  In
both cases, the rise time is similar to the observed flare from Oct 9,
indicating that perhaps strong flares from this source are common, and
have similar profiles. 

It is therefore possible that the persistent emission is simply a
superposition of several, possibly weaker, flares.  If this is the
case, the flare duty cycles for these two sources approach 100\%.
This is similar to LHS\,2065, in which Mart{\' i}n \& Ardila
(2001)\nocite{ma01} find that weak H$\alpha$ flares may be more common
than strong flares (see \S\ref{sec:ss}).  The implications of such
sustained flaring emission are discussed in \S\ref{sec:disc}.

We find no evidence for flares from the other
nine sources.  However, for most of the sources this is not surprising
given the observed duty cycles of strong flares, and the required
luminosities for a significant detection.  The most luminous flare in
the sample, from \tvlms{}, can only be detected out to a distance of
$\approx 20$ pc with significance $\sim 5\sigma$.  Five of the nine
undetected sources lie at or beyond this distance.  For the other four
sources the flares could be weaker, undetected in the short
observations ($\approx 2$ hr), or simply do not exist.  Therefore, the
lack of radio emission does not significantly constrain the production
of flares in these sources, and the upper limits only provide
constraints on the persistent emission.

\begin{figure} 
\plotone{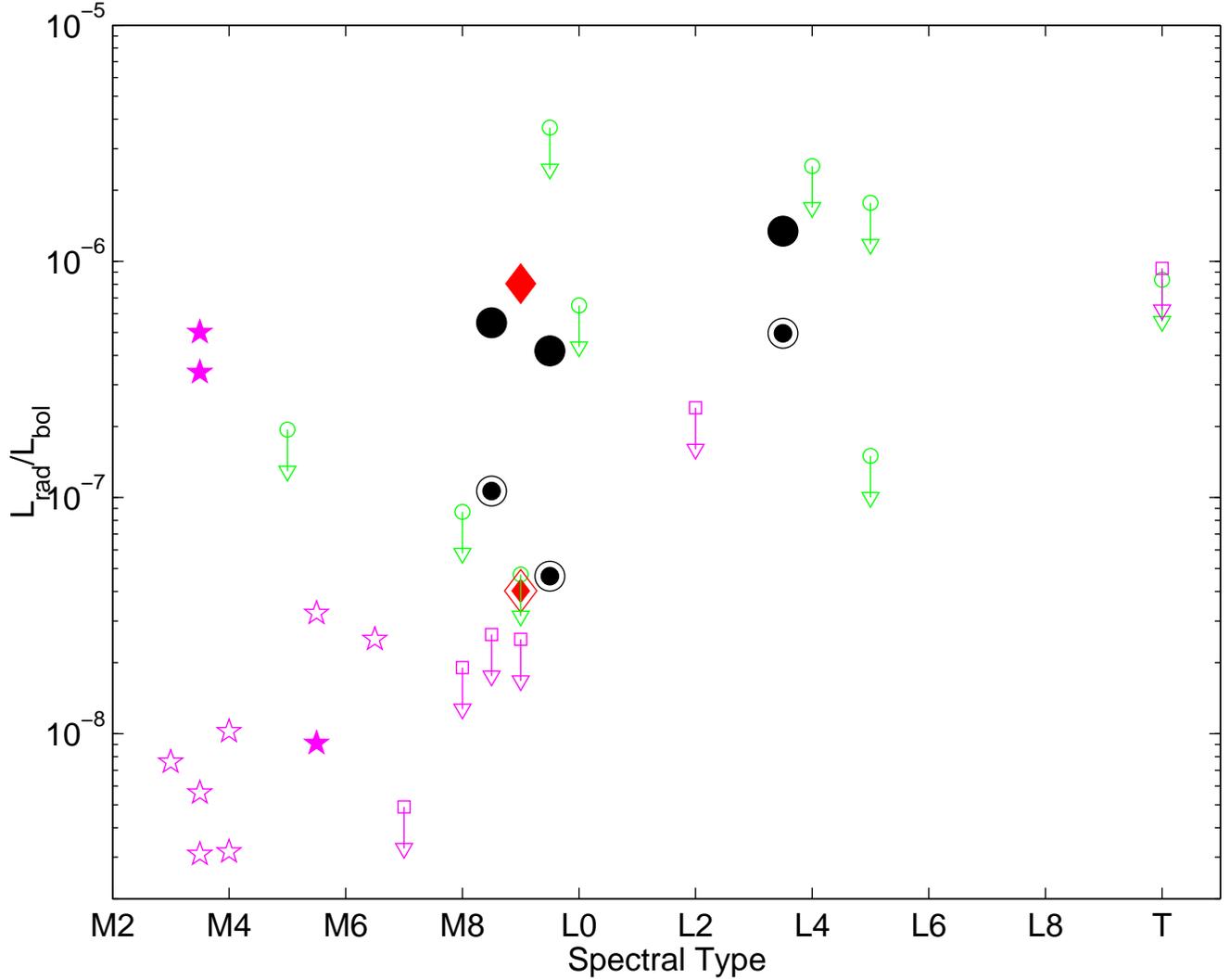}
\caption[]{Ratio of flare luminosities (filled symbols) and persistent 
luminosities (open symbols and filled/open symbols) to bolometric 
luminosities vs. spectral type.  Sources from this survey are designated 
by circles, \lp{} (B01) is designated by a diamond, the sources from 
Krishnamurthi, Leto \& Linsky (1999) are designated by squares, and 
sources in the range M3-M6.5 (Cash, Charles \& Johnson 1980;
Pallavicini, Wilosn \& Lang 1985; Leto et al. 2000; Stepanov et
al. 2001) are designated by stars.  The upper limits represent $3\sigma$ 
levels.  When the bolometric luminosity of the source
is not known, we estimated the value using other sources with the same
spectral type (see Table~\ref{tab:sources}).  While the sample size is
still small, it appears that there is no decline in
radio activity between spectral types M3--L3.5, indicating that the
radio and H$\alpha$ emission are possibly decoupled.  
\label{fig:lumvsspec1}}
\end{figure}

The strong quiescent emission in \tvlms{} and 2MASS\,0036+18 can only
be detected out to a distance of $\sim 15$
pc in similar observations.  Thus, some of the non-detections do not
provide significant constraints given the detected luminosities.  In
Figures~\ref{fig:lumvsspec1} and \ref{fig:lumvsspec2}, and
\S\ref{sec:disc} we show that most of the upper limits on the ratio of
radio luminosity to bolometric luminosity are consistent with the
persistent luminosities of the detected sources.  The most notable
exception is the upper limit on emission from LHS\,2065.  We see from
Figures~\ref{fig:lumvsspec1} and \ref{fig:lumvsspec2} that the upper
limit is comparable to the detection of quiescent emission from
LP944-20 and the possible quiescent emission from \bris{}.  This
non-detection also contrasts with the detection of H$\alpha$ emission
from this source (Mart{\' i}n \& Ardila 2001)\nocite{ma01}.  In
addition, a comparison of the upper limits from Krishnamurthi, Leto \&
Linsky (1999)\nocite{kal+99} to the detections in this survey
provides significant constraints on the lack of emission
from their M7-M9 objects.  As shown in Figures~\ref{fig:lumvsspec1}
and \ref{fig:lumvsspec2}, their upper limits are up to an order of
magnitude fainter in $L_{\rm rad}/L_{\rm bol}$ than the detected
sources in this survey.  We discuss the implications of these
non-detections in detail in \S\ref{sec:disc}.

\section{Analysis of the Radio Emission}
\label{sec:flare}

To derive the properties of the flares and persistent emission, and
the physical conditions that give rise to this radio emission, we model
the lightcurves of \tvlms{} and \bris{} with a Gaussian profile:
\begin{equation}
F_\nu(t)=F_{\nu,{\rm q}}+F_{\nu,0}{\rm
exp}\Bigl[-\frac{1}{2}\Bigl(\frac{t-t_0}{\sigma_{\rm
t}}\Bigr)^2\Bigr]. 
\label{eqn:model}
\end{equation}
Here, $F_{\nu,{\rm q}}$ is the quiescent component, $t_0$ is the flare
peak time relative to the start of the observation, $\sigma_{\rm t}$
is the HWHM duration of the flare, $F_{\nu,0}$ is the peak flux, and
we absorbed the normalization factor of $1/\sqrt{2\pi\sigma_{\rm t}^2}$
into the definition of $F_{\nu,0}$.  The lightcurve of 2MASS\,0036+18
from Oct 9 is not well-described by this model, and we instead
use an exponential profile, in which the Gaussian in
Equation~\ref{eqn:model} is replaced by rising and decaying
exponentials with timescales $\tau_{\rm r}$ and $\tau_{\rm d}$,
respectively (Figure~\ref{fig:mass2}).  In Table~\ref{tab:prop} we
summarize the derived parameters for each source.  We note that the
profile of the flare from 2MASS\,0036+18 is somewhat unusual since
flares usually exhibit a rapid rise and exponential decay as opposed
to the slow rise and fast decay in this case (Figure~\ref{fig:mass2}).
A similar profile was found for two of the three flares from \lp{}
(B01). 

The fits to the lightcurves of \tvlms{} (Figure~\ref{fig:tvlm}) and
2MASS\,0036+18 (Figure~\ref{fig:mass2}), have unsatisfactory $\chi^2$
values, $\sim 3$ per degree of freedom.  This is due to significant
variability in the persistent component, indicating that it is
possibly a superposition of multiple flares, rather than a truly
constant component.

The statistical significance of the flare detections are determined by
the level at which we can rule out the hypothesis that $F_{\nu,0}=0$
in Equation~\ref{eqn:model}, coupled with the number of time bins in
each lightcurve.  For the flares
from \tvlms{} and 2MASS\,0036+18 the detections are secure, with
significance levels of $\sim 20\sigma$.  The flare from \bris{} has a
lower statistical significance of $4.5\sigma$, but this is still
high enough for a significant detection.  We further tested the
significance of this flare by constructing equivalent lightcurves for 
random positions on the map.  We find an average fluctuation of
$2\sigma$, and no fluctuations larger than $2.5\sigma$, indicating
that the increase in flux in \bris{} is a likely flare.  

It is important to note that even if the short-timescale increase in
flux in the \bris{} lightcurve is simply a
$4.5\sigma$ statistical fluctuation in the lightcurve, the flux
density of this source from the combined observations on Jan 25 and
May 21 is $83\pm 15$ $\mu$Jy, and hence significant at the $5.5\sigma$
level.  Therefore, whether this detection is of persistent emission or
a faint flare, this source is radio active.

The peak-flare luminosities are in the range $L_{\nu,{\rm f}}\approx
(6-13)\times 10^{13}$ erg sec$^{-1}$ Hz$^{-1}$, more luminous
than the brightest Solar flares, $L_{\nu,{\rm f}}\approx 2\times
10^{12}$ erg sec$^{-1}$ Hz$^{-1}$ (Bastian, Benz \& Gary
1998)\nocite{bbg98}, but somewhat weaker than flares from active
early M dwarfs (e.g. White, Jackson \& Kundu 1989)\nocite{wjk89}.
The ratio of flaring radio to bolometric luminosity is higher
by a factor of $\sim 10^5$ than for Solar flares, but is very similar
to flares on early M dwarfs (see Figures~\ref{fig:lumvsspec1} and
\ref{fig:lumvsspec2}) since the lower luminosity of the flares
relative to those from early M dwarfs is compensated by the lower
bolometric luminosity of the late M and L dwarfs in this sample.
The persistent luminosities are lower by factors of 3, 5, and 9 than
the flare luminosities in 2MASS\,0036+18,
\tvlms{}, and \bris{}, respectively.  

The radio band energy release in the flares is $E_R\sim 1.4\times
10^{23}d^2_{\rm pc}\nu_{\rm GHz}F_{\nu,{\rm mJy}}\sigma_{t,{\rm
min}}\approx (2-10)\times 10^{26}$ erg, with the lowest energy release
in the flare from \bris{}, and the highest in the flare from \tvlms{}. 
In making this estimate we have assumed that the flare spectral energy
distribution peaks near 8.5 GHz, and has steep spectral slopes below
and above this frequency, so that the bulk of the energy release is at
$\sim 8.5$ GHz.  This assumption is justified since for
synchrotron emission (see \S\ref{sec:mag}) the spectral slope below
the peak (self-absorbed portion) is $F_\nu\propto \nu^{2.5}$, while
above the peak typical values ranges from $\sim \nu^{-1.5}$ to
$\nu^{-4}$ (e.g. Dulk \& Marsh 1982)\nocite{dm82}.  The high fraction
of circular polarization in the two bright flares from \tvlms{} 
and \masss{} indicates that the
observing frequency (8.46 GHz) lies at or above the peak frequency,
while observations of flares from several early M dwarfs (Leto et
al. 2000)\nocite{lpl+00} show that in these sources the flux drops 
significantly above
8.5 GHz.  In fact, for the M5.5 dwarf UV Cet the spectral slope above
8.5 GHz is constrained to be lower than $-2.3$.  Therefore, it is likely 
that most of the energy in the observed flares in this survey is
released at 8.5 GHz.

\subsection{Violation of the Guedel--Benz relations}
\label{sec:gb}

One of the puzzling facts about the radio emission from \lp{} (B01)
is that despite its spectral type (M9) it
violates the Guedel--Benz relations (which hold for spectral types
$<$M7) by four orders of magnitude.  Here we examine whether any of
the sources detected in this survey show a similar violation of these
relations. 

Based on these relations, and the flare duty cycle we find that the
expected X-ray luminosity of \bris{} is $L_{\rm X,pred}\approx
2.6\times 10^{28}$ erg sec$^{-1}$, dominated by the persistent
component.  This value is $\gtrsim 10^3$ times larger than an upper
limit of $L_{\rm X}<2.6\times 10^{25}$ erg sec$^{-1}$ from a 63.2 ksec
ROSAT observation (Neuh{\"a}user et al. 1999)\nocite{nbc+99}.  Even if
we neglect the persistent emission, the predicted  luminosity due to
flares is a factor of $>150$ too high.  Finally, if the observed
increase in the radio flux is a statistical fluctuation rather than a genuine
flare, and the flux of the source is steady at $83\pm 15$ $\mu$Jy,
this emission violates the Guedel--Benz relations by a factor of
$>1700$.  Therefore, regardless of whether the radio emission from this
source is flaring or persistent (or both), it violates the
Guedel--Benz relations by several orders of magnitude.  This is
puzzling since, as in the case of \lp{}, \bris{} has spectral type
M9.5, indicating that radio and X-ray emission become uncorrelated
over a narrow range of spectral type.

As far as we know, \tvlms{} and 2MASS\,0036+18 have not been observed
in X-rays, so a similar analysis is not possible.  However, we can
still show that the radio emission from these sources is much stronger
relative to the bolometric luminosity than in early-mid M dwarfs, 
based on a comparison of the predicted peak X-ray luminosity and the
bolometric luminosity.  In the case of \tvlms{}, the predicted peak
X-ray luminosity is $\approx 4\times 10^{29}$ erg sec$^{-1}$, which is
approximately 25\% of the bolometric luminosity of this source.  In
2MASS\,0036+18 this ratio is even higher, $\sim 50\%$.  Liebert et
al. (1999)\nocite{lkr+99}, among others, have found a similar result
for H$\alpha$ flares, and noted that these flares may release more
energy relative to the bolometric luminosity than in earlier spectral
types.  

Therefore, while the Guedel--Benz relations hold for spectral types
$<$M7, there is some indication from \lp{} and \bris{} that these
relations break down around spectral type M9.  This possibly indicates
a sudden change in the mechanism that gives rise to the radio
emission.  We note that radio and X-ray observations of a larger
sample are required to test this idea comprehensively.

\subsection{Magnetic Fields and Coronal Densities}
\label{sec:mag}

So far, we have analyzed the radio emission in a relatively
model-independent way.  However, the next step is to determine the 
emission mechanism and physical conditions that gave rise to the
flares and persistent emission.  The first step in this analysis is to
calculate the brightness temperatures: 
\begin{equation}
T_{\rm b}\approx 2\times 10^9F_{\rm \nu,mJy}d_{\rm
pc}^2\nu_{\rm GHz}^{-2}(R/R_{\rm J})^{-2}.
\end{equation}
Assuming that the emission is coronal, we can use  $R\sim
(2-4)R_{\rm s}\sim (2-4)\thinspace R_{\rm J}$ as the size of the
corona in units of Jupiter radii (Linsky \& Gary 1983; Dorman, 
Nelson, \& Chau 1989; Burrows, Hubbard, \& Lunine 1989; Chabrier 
et al. 2000; Leto et al. 2000)\nocite{cba+00,lpl+00,bhl89,dnc89,lg83}.
These size estimates lead to brightness temperatures of $\approx
10^{8}-10^{9}$ K during the flares, and $\approx 10^7-10^8$ K for the
persistent emission.  Alternatively, it is possible that the flares
arise from much smaller regions (e.g. coronal loops), in which case
the brightness temperatures would exceed $10^{11}$ K, if $R_{\rm
loop}\sim 0.1R_{\rm s}\sim 0.1R_{\rm J}$.  While we have no data to
constrain the size of the coronae in these sources, it is likely that
the actual brightness temperatures are within the range defined by
these estimates.

The high brightness temperatures indicate that the emission is
non-thermal.  Most likely it arises from the incoherent
gyrosynchrotron or synchrotron processes (Dulk \& Marsh 1982; Dulk
1985)\nocite{dm82,dul85}, but coherent emission is also applicable if
the flares actually have $T_{\rm b}\sim 10^{11}$ K.  The high degree
of circular polarization in the flares from \tvlms{} and
2MASS\,0036+18 possibly favors coherent emission, but can also be
achieved in gyrosynchrotron emission (Dulk 1987)\nocite{dul87}.   On
the other hand, the
high frequency of the observed emission, $\nu=8.46$ GHz, favors
incoherent emission since coherent emission is limited to $\lesssim 
10$ GHz (Bastian, Benz \& Gary 1998; Stepanov et al. 2001).  Since it 
is not clear which emission process produces the observed radio emission,
we estimate the magnetic field strengths and coronal densities using
both.  

To accurately calculate the magnetic field strengths and coronal
densities in the radio active sources using the gyrosynchrotron 
formulation,
requires detailed spectral and geometrical information, since the
results depend sensitively on the high-frequency spectral slope, the
viewing angle, and the geometry of the emission region (Dulk \& Marsh
1982; Dulk 1985)\nocite{dm82,dul85}.  For approximate values we can
instead use the synchrotron formulation, along with the
approximation $f_{\rm circ}\approx 3/\gamma_{\rm min}$, where
$\gamma_{\rm min}$ is the minimum Lorentz factor of the radio-emitting
electron distribution.  For \tvlms{} and 2MASS\,0036+18, the Lorentz
factor is $\gamma_{\rm min}\approx 5$, while for \bris{} it is
$\gamma_{\rm min}\sim 10$.
The resulting magnetic field strengths are $B\approx 57\nu_{\rm
m,GHz}\gamma_{\rm min}^{-2}\approx 20$ G in \tvlms{} and
2MASS\,0036+18, and $\approx 5$ G in \bris{}.  These values should be
treated with caution, and they can vary upward by an order of
magnitude or more.  In fact, if we choose characteristic values for
the radio spectral index and the viewing angle, we find
gyrosynchrotron values of $B\sim 350$ G for \tvlms{} and
2MASS\,0036+18, and $B\sim 50$ G for \bris{}.  

For comparison, the Solar magnetic field as inferred from a
multi-wavelength study of three Solar flares (Kundu et
al. 2001)\nocite{knw+01} ranges from $300-900$ G at the loop footpoints,
while fields in active early M dwarfs reach strengths of a few kG
(Saar \& Linsky 1985; Haisch, Strong \& Rodono 1991; Johns-Krull \&
Valenti 1996; Stepanov et al. 2001)\nocite{skz+01,jkv96,hsr91,sl85}.   

The total number of electrons in the synchrotron formulation is easily
calculated using $N_{\rm e}\approx 10^{36}d_{\rm pc}^{2}F_{\nu a,{\rm
mJy}}\gamma_{\rm min}^{-2}B_{\rm min}^{-2}\approx (5-20)\times
10^{33}$.  However, to calculate the densities we have to know the
geometry of the flare region.
In the gyrosynchrotron formulation, using the magnetic field values
quoted above, we find column densities of $\sim 10^{19}$ cm$^{-2}$ for
\tvlms{} and 2MASS\,0036+18, and $\sim 10^{22}$ cm$^{-2}$ for \bris{}.
These densities are similar to those calculated for Solar and M dwarf
flares (Bruggmann \& Magun 1990; Stepanov et
al. 2001)\nocite{bm90,skz+01}.  

Alternatively, if the flare brightness temperatures are of the order
of $10^{11}$ K, then we must use coherent emission to
estimate the magnetic field strengths and the densities.  There are
two main coherent processes, electron cyclotron maser (ECM), and
plasma emission, which operate at the fundamental frequencies
$\nu_c\approx 2.8B$ MHz and $\nu_p\approx 9n^{1/2}$ kHz, respectively.
Recently, Stepanov et al. (2001)\nocite{skz+01} have shown for the
flare star AD Leonis that emission due to the ECM process is strongly
damped by thermal electrons in the corona.  They conclude that for
typical coronal temperatures and densities plasma emission is the more
likely process. 

Plasma radiation is primarily emitted at $\nu\approx \nu_p$.
Thus, we can simply use $\nu_p\sim 8.5$ GHz to find $n\sim 10^{12}$
cm$^{-3}$, and approximate $\nu_c\lesssim 0.5\nu_p$ to find $B\lesssim
1500$ G.  These values are consistent with the
gyrosynchrotron-derived values.  Therefore, we conclude that
regardless of the exact emission mechanism or magnetic field
configuration, the observations are consistent with magnetic fields in
the range $10-10^3$ G, and coronal densities of the order of $10^{12}$
cm$^{-3}$.  We discuss the implications of the inferred magnetic field
values and coronal densities in more detail in \S\ref{sec:disc}.

\subsection{Heating Mechanisms}
\label{sec:heat}

We next investigate whether our radio data support a particular model
of coronal particle acceleration.  Heating can be due to acoustic
waves generated by the same turbulent convection that presumably gives
rise to the magnetic fields  (Bohn 1984; Ulmschneider, Theurer, \&
Musielak 1996)\nocite{b84,utm96}.  Ulmschneider et al. (1996) provide
an estimate of the total flux due to acoustic heating, $F_{\rm A}$,
as a function of effective temperature and surface gravity based on an
extended Kolmogorov turbulent energy spectrum.  We use their Table 1
to estimate ${\rm log}(L_{\rm A}/L_{\rm bol})\sim -13$ for our survey 
sources.  In comparison, the values for the observed radio flares are 
${\rm log}(L_{\rm rad,f}/L_{\rm
bol})\sim -6$ (Figures~\ref{fig:lumvsspec1} and \ref{fig:lumvsspec2}).
Therefore acoustic 
heating cannot supply the necessary energy to power the flares, and
similarly, cannot explain the persistent emission with ${\rm
log}(L_{\rm rad,q}/L_{\rm bol})\sim -7$. 

A more plausible heating mechanism is magnetic reconnection
(Bastian, Benz \& Gary 1998; Sturrock 1999)\nocite{bbg98,s99}.  This
model can also explain the persistent emission if it is a
superposition of weaker flares.  In the context of this model, the
flares are produced when coronal magnetic loops reconnect, release
energy, and create a current sheet along which ambient electrons are
accelerated.   Recent work on three-dimensional reconnection shows
that the particles are accelerated into a power-law energy
distribution (Schopper, Birk \& Lesch 1999)\nocite{sbl01}, as required
for the synchrotron and gyrosynchrotron discussion in \S\ref{sec:mag}.
The accelerated electrons drive an outflow of hot plasma into the
corona as they interact with, and heat chromospheric material
(i.e. chromospheric evaporation).  The interaction of the outflowing
plasma with the same electons produces X-ray emission via the
bremsstrahlung process (Neupert 1968; Hawley et
al. 1995; Guedel et al. 1996)\nocite{gbs+96,hfs+95,neu68}.  This
so-called Neupert effect indicates that there is a {\it causal}
connection between particle acceleration, which is the source of radio
emission, and plasma heating, which results in X-ray emission.  This 
connection explains the phenomenological Guedel--Benz relations. 

It is possible that the violation of the Guedel--Benz relations in the
emission from \lp{} and \bris{} indicates that the Neupert effect no
longer holds in late M dwarfs.  A direct test of this possibility
requires simultaneous observations of such sources in the radio and
X-rays.  Future surveys of activity in late M and L dwarfs would
be considerably more effective if they followed such a strategy.

It is also possible that the energy source for the radio emission in
the active survey sources and \lp{} is an altogether different
process, in which case a causal relation between
radio emission and activity in other bands would not be required.  At 
present it is not
possible to distinguish between these different possibilities, but
future observations with wide spectral coverage will likely provide an
answer.

\section{Discussion}
\label{sec:disc}

Before we proceed to interpret the results of this survey
in the context of correlations between radio activity and
physical source parameters, we summarize the main observational
results: (i) the observed flares have approximately equal
luminosity, ratio of radio to bolometric luminosity, and
released energy, (ii) the persistent emission has approximately the same
luminosity, and it appears to be variable, (iii) LHS\,2065 has an
upper limit on the ratio of radio to bolometric luminosity which is
lower than the detected persistent emission, despite being active in
H$\alpha$, (iv) the emission from \bris{}, as well as \lp{}, violates
the Guedel--Benz relations by several orders of magnitude, (v)
the inferred magnetic field strengths are $\sim 10-10^3$ G, and the
coronal densities are $\sim 10^{12}$ cm$^{-3}$, and (vi) the heating
mechanism is probably magnetic reconnection, but it is possible that
the Neupert effect is violated.

\begin{figure} 
\epsscale{0.8}
\plotone{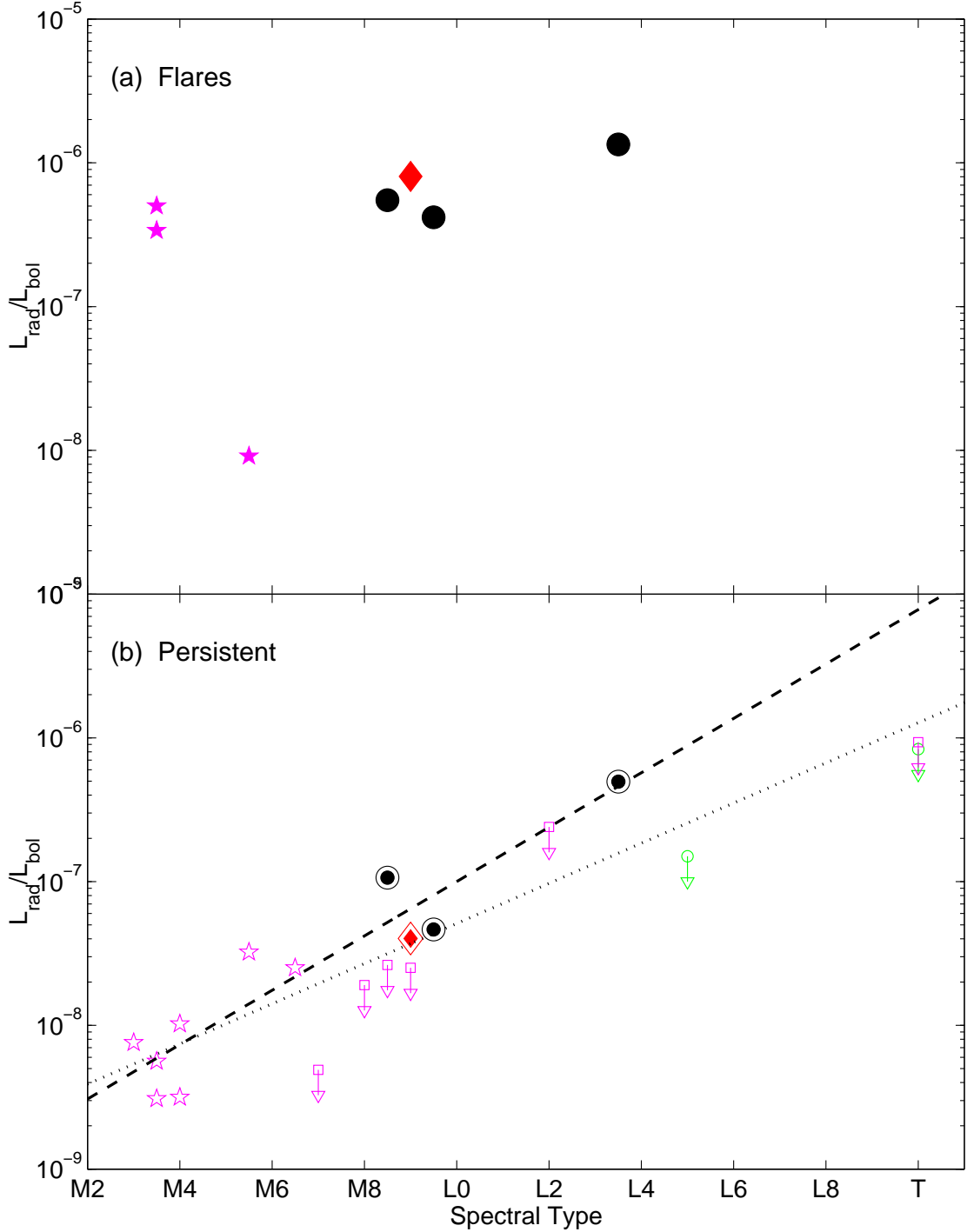}
\caption[]{(a) Ratio of radio to bolometric luminosities for detected
flares from M and L dwarfs.  Symbols are as in
Figure~\ref{fig:lumvsspec1}.  With the exception of Proxima Centauri
(M5.5), the radio activity appears to cluster around ${\rm log}(L_{\rm
rad,f}/L_{\rm bol})\approx -6$.  This possibly indicates a saturation
effect.  (b) Ratio of radio to bolometric luminosities for persistent
emission from the detected sources, and upper limits which are similar
to or lower than these detections.  The dashed line is a linear fit to
the detections, while the dotted line includes the upper limits.  In
both  cases there is an increase in ${\rm log}(L_{\rm rad,q}/L_{\rm
bol})$  with spectral type.  The same ratio for H$\alpha$ emission
drops significantly beyond M7.  These observations indicate that the
radio and H$\alpha$ emission are probably uncorrelated at the bottom
of the main sequence.  The upper limits that violate the relation, and
have measured rotational velocities, have $v{\rm sin}i<10$ km
sec$^{-1}$.  This result is discussed in detail in \S\ref{sec:disc}
and Figure~\ref{fig:lumvsrot}.
\label{fig:lumvsspec2}} 
\end{figure}

A comparison of these results to the radio properties of \lp{}
(B01) is illustrative.  The peak luminosity of the flares from all
four sources varies by only a factor of two, and the energy release
only varies by a factor of five.  The ratio of the flare peak
luminosity to the bolometric luminosity of each source is ${\rm
log}(L_{\rm rad,f}/L_{\rm bol})\approx -6$, and the ratio for the
quiescent luminosities is ${\rm log}(L_{\rm rad,q}/L_{\rm bol})\approx
-7.5$ to $-6.5$ (Figures~\ref{fig:lumvsspec1} and
\ref{fig:lumvsspec2}).  

We find similar results when we compare this sample to other M dwarfs
that have been observed in the radio.  For example, flares have been
detected from the dM3.5e dwarf Ad Leo (Stepanov et
al. 2001)\nocite{skz+01}, the dM4e dwarf Rst 137B (Lim
1993)\nocite{lim93}, and the dM5.5e dwarf Proxima Centauri (Lim,
White \& Slee 1996)\nocite{lws96}, with ratios of $L_{\rm rad}/L_{\rm
bol}$ that are in the range of $\sim -6$ to $-8$.  These values are
plotted in Figures~\ref{fig:lumvsspec1} and \ref{fig:lumvsspec2}a,
and we can see that with the exception of the flare from Proxima
Centauri, which is fainter by approximately two orders of magnitude,
these values are similar to those observed in this survey and in
\lp{}.  The flat distribution of $L_{\rm rad}/L_{\rm bol}$ for the
flares from the sources with spectral types M8-L3.5, and perhaps
M3.5-L3.5 (with the exception of Proxima Centauri), possibly
points to a ``saturation'' effect, in which $L_{\rm rad,f}/L_{\rm bol}$
reaches a maximum value that is independent of spectral type or other
physical properties.  It 
is beyond the scope of this paper and data to assess this possibility in 
detail.

The pattern of persistent radio emission in early M dwarfs through L
dwarfs is more puzzling.  In Figures~\ref{fig:lumvsspec1} and
\ref{fig:lumvsspec2}b we plot the ratios of persistent radio to
bolometric luminosities for the detections and upper limits in this
survey, the upper limits from Krishnamurthi, Leto \& Linsky (1999), and
detections in the range M3-M6.5 (Cash, Charles \& Johnson 1980;
Pallavicini, Wilosn \& Lang 1985; Leto et
al. 2000)\nocite{lpl+00,pwl85,ccj80}. 
Despite the small sample, there is a clear increase in the ratio of
persistent radio to bolometric luminosity over the range M3-L3.5
(dashed and dotted lines in Figure~\ref{fig:lumvsspec2}b).  This increase is 
in direct contradiction with the steep decline in H$\alpha$ activity
beyond spectral type M7 (Gizis et al. 2000).  In fact, while the
H$\alpha$ activity drops by approximately 1.5 orders of magnitude
between M7 and L3.5, the radio activity appears to {\it increase} by a 
similar amount in the same spectral range.  This indicates that the radio 
and H$\alpha$ activity are probably not correlated at the bottom of the
main sequence.

\begin{figure}
\epsscale{1.0} 
\plotone{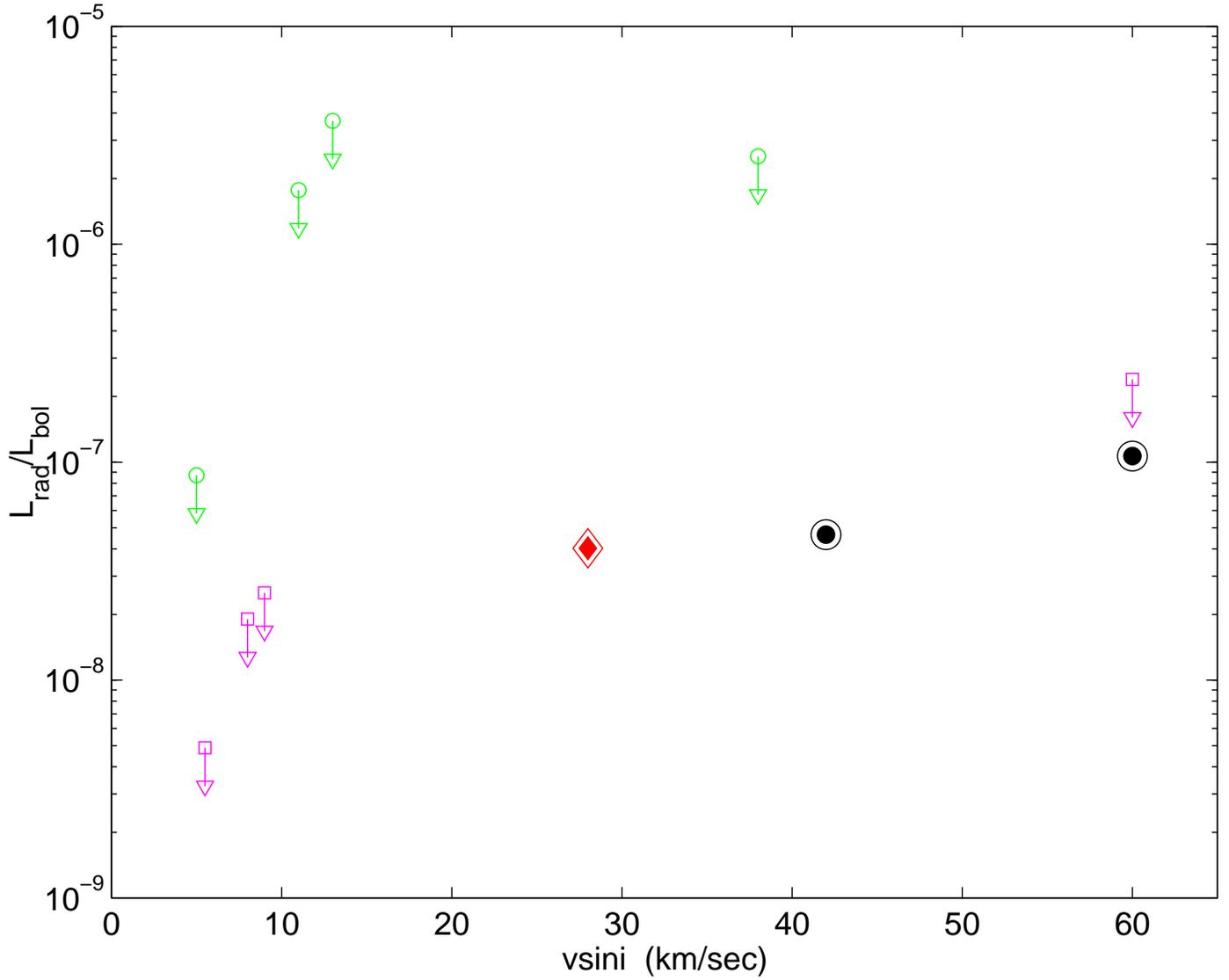}
\caption[]{Ratios of persistent radio luminosity to bolometric
luminosity for sources with measured rotational velocities.  Symbols
are as in Figure~\ref{fig:lumvsspec1}.  The sources with detected
radio emission have $v{\rm sin}i>30$ km sec$^{-1}$, while sources 
which have upper limits lower than the detected levels have $v{\rm
sin}i<10$ km sec$^{-1}$.  This trend is based on a small number of
objects, but it provides tantalizing evidence for a correlation
between radio activity and rotation. 
\label{fig:lumvsrot}}
\end{figure}

Another interesting result from Figure~\ref{fig:lumvsspec2}b is that
there are several upper limits, which appear to fall below the apparent
steady increase in $L_{\rm rad}/L_{\rm bol}$.  These upper limits are
for 2MASS\,1507$-$16 (L5) and \sdsss{} (T6) from this survey, and five
sources from Krishnamurthi, Leto \& Linsky (1999): VB\,8 (M7), VB\,10 (M8),
GL\,569B (M8.5), LHS\,2065 (M9), and Gl\,229B (T); we use the upper
limit for LHS\,2065 from 
Krishnamurthi, Leto \& Linsky (1999) since it is approximately 3 times lower
than the value measured in our survey.  While some variance in
activity levels within the sample of M and L dwarfs is not unexpected,
it is interesting to investigate whether these source have some
property which differentiates them from the detected sources.  In 
Figure~\ref{fig:lumvsrot} we plot the ratio of persistent radio to
bolometric luminosity as a function of the source rotational velocity,
and we find that of these seven sources the three with measured
rotational velocities have $v{\rm sin}i<10$ km sec$^{-1}$, while the
detected sources have $v{\rm sin}i>30$ km sec$^{-1}$.  In fact, the
source with the most constraining upper limit (VB\,8) has the lowest
rotational velocity in the sample, while the source with the highest value 
of $L_{\rm rad}/L_{\rm bol}$ (\tvlms{}) has the highest rotational velocity.  
The upper limit on the radio
activity from the fast rotator Kelu\,1 (Krishnamurthi, Leto \& Linsky 1999),
with $v{\rm sin}i\approx 60$ km sec$^{-1}$ (Basri et
al. 2000)\nocite{bma+00} is three times higher than the detected fast
rotators, and therefore does not constrain the possible relation
between radio activity and rotational velocity.

We note that while these sources provide tantalizing evidence for a
relation between rapid rotation and radio activity, the sample is
still very small.  Future observations of a larger sample will provide
a better test of this possible relation.  However, it is already
possible to rule out the idea that high rotational velocities quench
the magnetic dynamo through supersaturation (Randich
1998)\nocite{ran98}, as was inferred from the drop in persistent
H$\alpha$ and X-ray activity in late M and L dwarfs. 

The most puzzling implication of the observed radio emission,
regardless of its possible correlation with fast rotation, is that
late M and L dwarfs posses appreciable and sustained magnetic
fields, as well as high particle densities in at least some regions of
their coronae.  This result seems to contradict recent calculations by
Mohanty et al. (2002)\nocite{mbs+02} which indicate that the generation
and propagation of magnetic energy decreases with decreasing effective
temperature, resulting in a decrease of coronal activity in late
spectral types.

In addition, the presence of magnetic fields and ionized
material indicates that magnetic braking should have slowed these
sources down considerably.  Thus, it is not clear how these sources
maintain such high rotational velocities.

No other correlations are apparent from the data.  For example,
the upper limit on the young object CRBR\,15 is approximately five
times higher than the faintest detected persistent emission, and
therefore precludes any conclusion on the role of age in the
production of radio emission.  Similarly, the upper limits on emission
from the confirmed binary systems \deniss{} and \gd{} are
approximately forty times higher than the detected sources, and it is
therefore not possible to assess the role of binary interaction in the
production of radio emission.

\section{Conclusions}
\label{sec:conc}

We detected flaring and persistent emission at 8.46 GHz from the M and
L dwarfs \tvlms{}, 2MASS\,0036+18, and \bris{}, but no emission from
the nine other survey sources.  These observations lead to several
interesting conclusions, which are puzzling given the decrease in
persistent H$\alpha$ and X-ray activity in late M and L dwarfs.  

The flares and persistent emission from the three survey sources, as
well as \lp{} (B01) have similar properties.  In particular, the
emission from these sources has similar $L_{\rm rad}/L_{\rm bol}$
values, and energy release.  This probably points to a common emission
mechanism and physical conditions (e.g. magnetic fields) in these
sources.  

In addition, while the dependence of H$\alpha$ activity on rotation
appears to break down in early-mid M dwarfs, the radio active sources
for which the rotational velocity is known are rapid rotators, with
the most luminous source in the sample, \tvlms{}, having the highest
rotational velocity, $v{\rm sin}i\approx 60$ km sec$^{-1}$.  At the
same time, the limits on radio activity in three slow rotators from
Krishnamurthi, Leto \& Linsky (1999) indicate that they are less
active than the detected sources in this sample, pointing to a
possible relation between rotation and radio activity.  Regardless of
whether a correlation actually exists, the radio emission requires
magnetic fields and coronal densities that should have resulted in
significant slow-down due to magnetic braking.  It is still not clear
how these sources maintain their rapid rotation.

Possibly related to the differences in radio activity and other
activity indicators is the violation of the Guedel--Benz relations in
\bris{} and \lp{}.  In these two sources, the predicted X-ray emission
is over-luminous relative to X-ray observations.  This indicates
that radio and X-ray activity are no longer correlated in some late M
dwarfs, in the same way that radio and H$\alpha$ activity appear to be
uncorrelated.  Moreover, it appears that the breakdown in the
Guedel--Benz relations occurs over a narrow spectral type range, M7-M9.

\begin{figure} 
\plotone{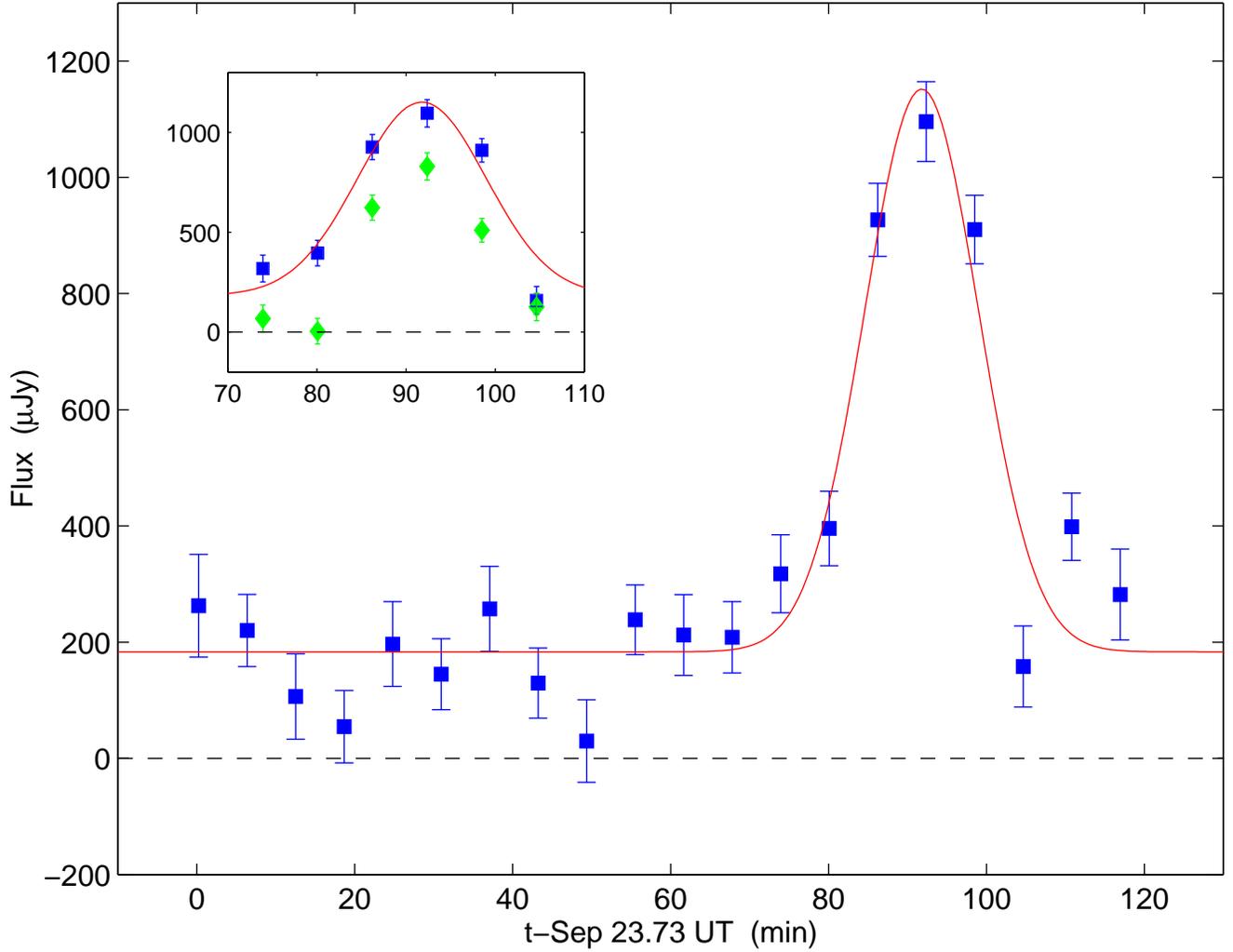}
\caption[]{Lightcurve of the 8.46 GHz emission from \tvlms{} on 2001,
Sep 23.73 UT, with 20 5.5-min bins.  We find a single flare and
persistent emission, which appears to be variable.  The solid line is
a Gaussian model (Equation~\ref{eqn:model}).  The inset shows the
circularly polarized flux (diamonds) and the total flux (squares).
The fraction of circular polarization near the peak of the flare is
$\approx 65\%$ (see Table~\ref{tab:prop}).
\label{fig:tvlm}}
\end{figure}

\begin{figure} 
\plotone{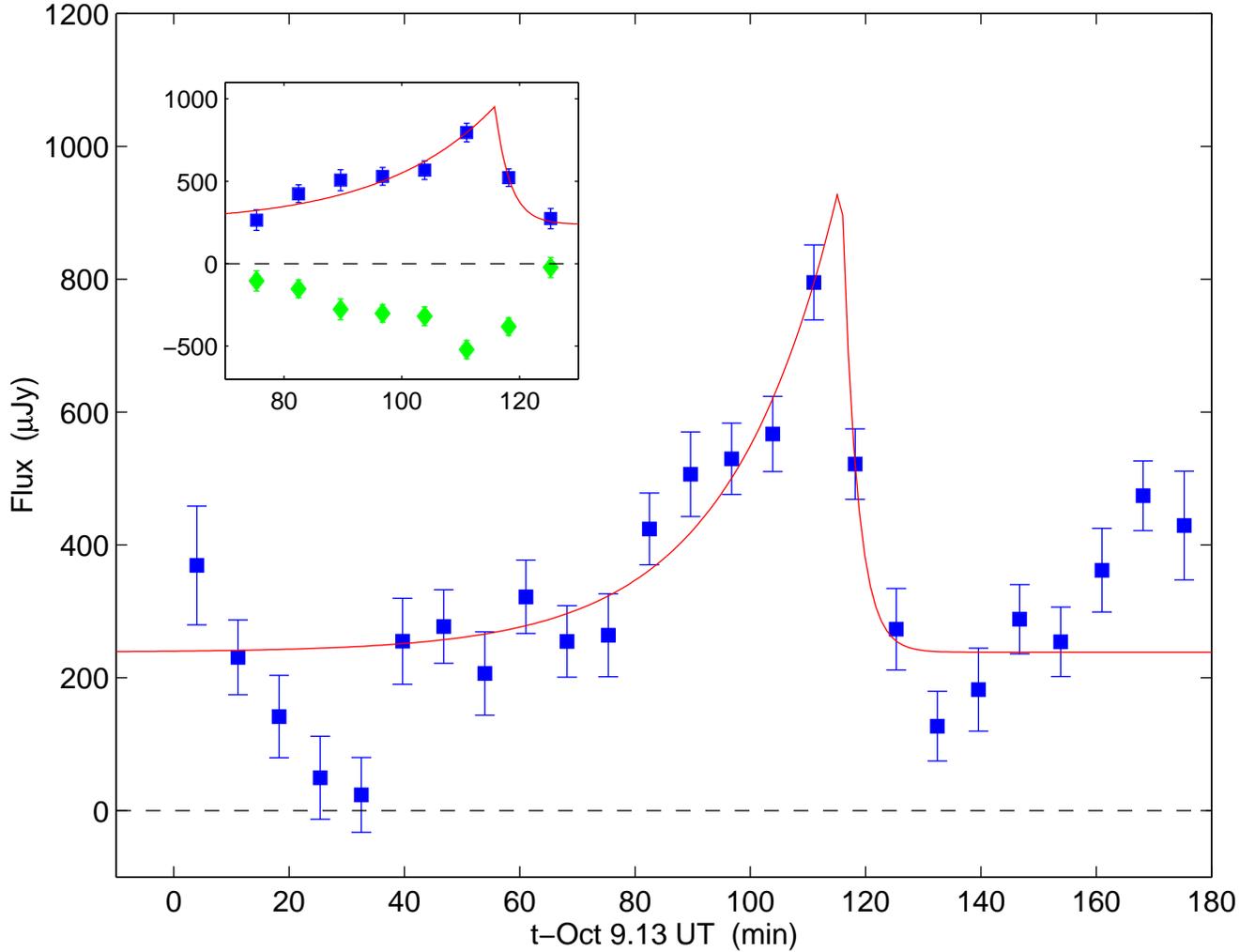}
\caption[]{Lightcurve of the 8.46 GHz emission from 2MASS\,0036+18 on
2001, Oct 9.13 UT, with 25 6.5-min bins.  We find a flare and
persistent emission, which appears to be strongly variable.  The steep
decline in flux during the first twenty minutes of the observation,
and the shallow rise during the last sixty minutes possibly signal two
additional strong flares.   The solid line is an exponential model.
The inset shows the circularly polarized flux (diamonds) and the total
flux (squares).  The fraction of circular polarization near the peak
of the flare is $\approx -65\%$.  The negative values indicate
left-handed circular polarization (see Table~\ref{tab:prop}).
\label{fig:mass2}}
\end{figure}

The observed persistent emission is clearly variable, possibly on
a timescale of the order of one hour (Figures~\ref{fig:tvlm} and
\ref{fig:mass2}).  It is possible that this emission is simply a
superposition of weak flares, in which case the activity duty cycle is
$\sim 100\%$.  However, if this component is truly persistent over
long timescales, it means that our understanding of magnetic field
generation in very low mass stars is far from complete.  Longer,
multi-frequency observations are required for a thorough understanding
of the origin of this emission component.

To address the possible correlation between rotation and radio activity, 
as well as the differences between the
activity indicators (i.e. radio, H$\alpha$, and X-rays), and to gain a
deeper understanding of physical processes in dwarf stars
it is necessary to carry out a comprehensive, multi-wavelength survey
of nearby M and L dwarfs.  In particular, it is crucial to observe a
wide variety of objects for each spectral type, and to reach ratios of
${\rm log}(L_{\rm rad}/L_{\rm bol})<-8$.  Despite the complexity of
such an effort, the results will impact a wide range of fields from
dynamo theory to the formation of brown dwarfs.

\acknowledgements  I thank B. Clarke for re-scheduling
some of the VLA time awarded to this project.  I also thank
R. Rutledge, D. Frail, S. Kulkarni, and D. Reichart for valuable
discussions, and the anonymous referee for many helpful comments and
suggestions.   This research has made use of the Online
Brown Dwarf Catalog (http://ganymede.nmsu.edu/crom/cat.html), which is
compiled and maintained by Christopher R. Gelino at New Mexico State
University, the SIMBAD database operated at CDS, Strasbourg, France,
and NASA's Astrophysics Data System Abstract Service

\clearpage
\begin{deluxetable}{lcccccccccc}
\tabcolsep0.04in\footnotesize
\tablewidth{0in}
\tablecaption{Properties of the Survey Sources \label{tab:sources}}
\tablehead {
\colhead {Source} &
\colhead {Spect.~Type.} &
\colhead {Dist.} &
\colhead {$T_{\rm eff}$} &
\colhead {Age} &
\colhead {$v{\rm sin}i$} &
\colhead {Log($L_{\rm bol}/L_\odot$)} & 
\colhead {Log($L_{{\rm H}\alpha}/L_{\rm bol}$)} & 
\colhead {Log($L_X$)} &
\colhead {Pred. $F_{\nu,R}$} &
\colhead {Flares?} \\
\colhead {} &
\colhead {} &
\colhead {(pc)} &
\colhead {(K)} &
\colhead {(Myr)} &
\colhead {(km/sec)} &
\colhead {} &
\colhead {} &
\colhead {(erg/sec)} &
\colhead {($\mu$Jy)} &
\colhead {} 
}
\startdata
\crbr & M5 & 160 & 2930 & $<3$ & & $-1.3$  & & $<28.0$  & $<0.1$ & no \\
LHS\,2243 & M8 & 17 & 2880 &  & $<5$ &  &  $-3.6$  &  & & yes \\
\tvlms & M8.5 & 10.5 & 2500/2880 &  & 60 & $-3.6$ &  &  &  & no \\
LHS\,2065 & M9 & 8.5 & 2100/2630 & $>500$ & 9 & $-3.5$ & $-4$ & 26.6 & 1.5 & yes \\
\bris & M9.5 & 12 & 2300 &  & $42\pm 8$ & $-3.5$  & $-6.3$  & $<25.4$ & $<0.05$ & yes \\
\pcs & M9.5 & 62 & 2000 & $<1000$ & $13\pm 3$ & $-3.7$ & $-3.4$ & $<27.2$ & $0.2$ & no \\
\masss & L0 & 27 & 2000 & $<1000$ & & $-3.6$ & $<-5.6$ & & & no \\
2MASS\,0036+18 & L3.5 & 9 & 1800 & $>1000$ &  &  &  &  &  & no \\
\gd & L4 & 32 & 1900 & $<1200$ & $38$ & $-4.0$  & $<-5.2$ &  &  & no \\
2MASS\,1507$-$16 & L5 & 8 & 1700 & $>1000$ &  &  &  &  &  & no \\
\deniss & L5 & 20 & $<1800$ & $<1000$ & $11$ & $-4.3$ & $<-5.3$ & $<27.3$ & $<3$ & no\\
\sdsss & T6 & 11 &  $\sim 900$ & & & $-5.3$ & $<-5.3$ & &  & no \\\hline
\lp & M9 & 5 & 2200 & $\sim 550$ & $30$ & $-3.8$ & $-5$ & $25.7$ & $0.5$ & yes 
\enddata
\tablecomments{List of objects observed in this survey; \lp{} is
included for completeness.  The columns are (left to right), (1) source
name, (2) spectral type, (3) distance, (4) effective temperature, (5)
approximate age, (6) rotational velocity, (7) bolometric luminosity,
(8) ratio of H$\alpha$ luminosity to bolometric luminosity; for \pc{}
the H$\alpha$ emission is persistent, while for LHS\,2243, LHS\, 2065,
\bris{}, and \lp{} it is flaring, (9) X-ray luminosity; for \lp{} this
is the peak-flare luminosity (Rutledge et al. 2000)\nocite{rbm+00},
(10) predicted radio flux
densities based on the X-ray luminosities and the Guedel--Benz
relations (Guedel \& Benz 1993; Benz \& Guedel 1994)\nocite{gb93,bg94}; these 
values are for 5 GHz and should be higher at most by a factor of four for the 
observing frequency in this survey, $\nu=8.46$ GHz, and (11) indicates whether 
H$\alpha$ or X-ray flares have been detected from the source.  A more detailed
description of the sources, and the relevant references are given in
\S\ref{sec:ss}}.   
\end{deluxetable}

\clearpage
\begin{deluxetable}{llccccc}
\tabcolsep0.05in
\tablewidth{0in}
\tablecaption{Very Large Array Observations \label{tab:vla}}
\tablehead {
\colhead {Source} &
\colhead {Date}      &
\colhead {Array Config.} &
\colhead {beam size} & 
\colhead {$t^{\rm on-source}$} &
\colhead {$\nu_0$} &
\colhead {S$\pm\sigma$} \\
\colhead {} &
\colhead {(UT)}      &
\colhead {} &
\colhead {(arcsec)} &
\colhead {(ksec)} &
\colhead {(GHz)} &
\colhead {($\mu$Jy)}
}
\startdata
\pc    & 2001 Jan 25.85 & A & $0.34\times 0.24$ & 3.7 & 8.46 & $<137$\\
\bri   & 2001 Jan 25.93 & A & $0.33\times 0.26$ & 5.0 & 8.46 & $84\pm 27$ \\
\mass  & 2001 Jan 26.01 & A & $0.30\times 0.26$ & 5.0 & 8.46 & $<125$ \\
\denis & 2001 Jan 29.34 & A$\rightarrow$B & $0.78\times 0.56$ & 4.9 & 8.46 & $<154$ \\
\sdss  & 2001 Jan 30.43 & A$\rightarrow$B & $0.65\times 0.37$ & 5.3 & 8.46 & $<106$ \\
\gd    & 2001 Jan 30.51 & A$\rightarrow$B & $0.64\times 0.31$ & 5.3 & 8.46 & $<97$ \\
\crbr  & 2001 Jan 30.60 & A$\rightarrow$B & $0.72\times 0.48$ & 4.5 & 8.46 & $<149$ \\
\denis & 2001 May 29.20 & CnB & $1.6\times 0.86$ & 4.3 & 8.46 & $<87$ \\
\bri   & 2001 May 21.54 & CnB & $4.4\times 2.9$ & 5.5 & 8.46 & $83\pm 18$ \\
\bri   & 2001 June 11.46 & CnB & $4.6\times 4.1$ & 6.4 & 4.86 & $<94$ \\
\mass  & 2001 June 11.58 & CnB & $3.0\times 1.7$ & 5.1 & 8.46 & $<88$ \\
\sdss  & 2001 June 20.24 & CnB & $4.7\times 1.9$ & 5.1 & 8.46 & $<68$ \\
\pc    & 2001 June 20.41 & CnB & $4.8\times 1.8$ & 5.0 & 8.46 & $<75$ \\ 
2MASSW J0036159+182110 & 2001 Sep 23.44 & DnC & $12.1\times 8.2$ & 7.8 & 8.46 & $135\pm 14$ \\ 
LHS\,2065 & 2001 Sep 23.55 & DnC & $12.5\times 7.5$ & 6.6 & 8.46 & $<81$ \\
LHS\,2243 & 2001 Sep 23.64 & DnC & $8.8\times 7.7$ & 6.6 & 8.46 & $<76$ \\
\tvlm & 2001 Sep 23.73 & DnC & $11.5\times 9.8$ & 6.6 & 8.46 & $308\pm 16$ \\
2MASSW J1507476$-$162738 & 2001 Sep 23.81 & DnC & $12.3\times 7.9$ & 7.8 & 8.46 & $<58$ \\
2MASSW J0036159+182110 & 2001 Oct 9.13 & DnC & $11.2\times 7.8$ & 9.6 & 8.46 & $327\pm 14$ \\
LHS\,2065 & 2001 Oct 12.66 & D & $11.2\times 7.8$ & 3.6 & 8.46 & $<95$ \\
LHS\,2243 & 2001 Nov 1.42 & D & $10.2\times 9.2$ & 11.1 & 8.46 & $<47$ \\
\enddata
\tablecomments{The columns are (left to right), (1) Source name, (2)
UT date of the start of each observation, (3) array configuration, (4) 
synthesized beam size, (5) total on-source observing time, (6)
observing frequency, and (7) peak flux density at the position of
each source, with the error given as the rms noise in the image; upper
limits are the flux at the position of the source plus $3\sigma$.}
\end{deluxetable}

\clearpage
\begin{deluxetable}{lcccc}
\tabcolsep0.05in\footnotesize
\tablewidth{0in}
\tablecaption{Properties of the Radio Emission \label{tab:prop}}
\tablehead {
\colhead {Parameter} &
\colhead {\tvlms{}} &
\colhead {2MASS\,0036+18 (Sep 23.44)} &
\colhead {2MASS\,0036+18 (Oct 9.13)} &
\colhead {\bris{} (May 21.54)}
}
\startdata
Ave. Flux ($\mu$Jy) & $308\pm 16$ & $135\pm 14$ & $327\pm 14$ & $83\pm 18$ \\
Ave. $f_{\rm circ}$ (\%) & $46\pm 4$ & $-46\pm 11$ & $-43\pm 3$ & $25\pm 17$ \\
$\chi^2_{\rm r}$ Const. & 21 & 1.5 & 7.2 & 2.2 \\ 
$\chi^2_{\rm r}$ Flare & 2.7 & --- & 2.9 & 1.2 \\
$F_{\nu,0}$ ($\mu$Jy) & $980\pm 40$ & --- & $720\pm 40$ & $360\pm 70$ \\
$\sigma_{\rm t}^{\rm flare}$ (min) & $13.8^{+1.0}_{-0.6}$ & --- &
$21.5^{+3.3}_{-2.7}$ & $5.6^{+2.0}_{-1.0}$ \\
$F_{\nu,{\rm q}}$ ($\mu$Jy) & $190\pm 15$ & --- & $240\pm 11$ & $25\pm 15$ \\
Peak $f_{\rm circ}$ (\%) & $66\pm 4$ & --- & $-62\pm 5$ & $30\pm 20$ \\
Energy (erg)        & $1.1\times 10^{27}$ & --- & $8.0\times 10^{26}$ & $2.2\times
10^{26}$
\enddata
\tablecomments{The rows are (top to bottom), (1) average flux density over
the entire observation, (2) average fraction of circular polarization
over the entire observation, (3) reduced $\chi^2$ assuming that each
lightcurve can be fit with only a constant term, (4) reduced $\chi^2$
when we include a flare, (5) flare peak flux
density, (6) flare FWHM duration, (7) quiescent flux density,
(8) fraction of circular polarization at the peak of the flare, and
(9) total energy release in the flare, assuming that the flares peaked
at 8.46 GHz.}
\end{deluxetable}

\clearpage 

\clearpage 

\clearpage 

\clearpage 

\clearpage 

\clearpage 

\clearpage 


\begin{thebibliography}{}

\bibitem[Basri \& Marcy(1995)]{bm95}
{Basri}, G. \& {Marcy}, G.~W. 1995, AJ, 109, 762.

\bibitem[Basri et al.(2000)]{bma+00}
Basri, G., et al. 2000, ApJ, 538, 363.

\bibitem[Bastian, Benz \& Gary(1998)]{bbg98}
{Bastian}, T.~S., {Benz}, A.~O., \& {Gary}, D.~E. 1998, ARAA, 36, 131.

\bibitem[Becklin \& Zuckerman(1988)]{bz88}
{Becklin}, E.~E. \& {Zuckerman}, B. 1988, Nature, 336, 656.

\bibitem[Benz \& Guedel(1994)]{bg94}
{Benz}, A.~O. \& {Guedel}, M. 1994, A\&A, 285, 621.

\bibitem[Berger et al.(2001)]{bbb+01}
{Berger}, E. et al.  2001, Nature, 410, 338.

\bibitem[Bessell(1991)]{bes91}
Bessell, M. S. 1991, AJ, 101, 662.

\bibitem[Bohn(1984)]{b84}
{Bohn}, H.~U. 1984, \aap, 136, 338.

\bibitem[Bruggmann \& Magun(1990)]{bm90}
{Bruggmann}, G., \& {Magun}, A. 1990, A\&A, 239, 347.

\bibitem[Burgasser et al.(2000)]{bkr+00}
{Burgasser}, A.~J., et al.  2000, AJ, 120, 473.

\bibitem[Burgasser et al.(2001)]{bkb+01}
Burgasser, A. J., et al. 2001, accepted to ApJ; astro-ph/0108452.

\bibitem[Burrows, Hubbard \& Lunine(1989)]{bhl89}
{Burrows}, A., {Hubbard}, W.~B., \& {Lunine}, J.~I. 1989, ApJ, 345, 939.

\bibitem[Burrows et al.(1997)]{bmh+97}
{Burrows}, A., et al. 1997, ApJ, 491, 856.

\bibitem[Cash, Charles \& Johnson(1980)]{ccj80}
{Cash}, W., Charles, P., \& Johnson, H. M.  1980, ApJ, 239, 23.

\bibitem[Chabrier et al.(2000)]{cba+00}
{Chabrier}, G., et al. 2000, ApJ, 542, 464. 

\bibitem[Delfosse, et al.(1997)]{dtf+97}
{Delfosse}, X., et al. 1997, A\&A, 327, L25.

\bibitem[Dorman, Nelson \& Chau(1989)]{dnc89}
{Dorman}, B., {Nelson}, L.~A., \& {Chau}, W.~Y. 1989, ApJ, 342, 1003.

\bibitem[Dulk \& Marsh(1982)]{dm82}
{Dulk}, G.~A., \& {Marsh}, K.~A. 1982, ApJ, 259, 350.

\bibitem[Dulk(1985)]{dul85}
{Dulk}, G.~A. 1985, ARAA, 23, 169.

\bibitem[Dulk(1987)]{dul87}
{Dulk}, G.~A. 1987, {\it Cool Stars, Stellar Systems, and the Sun.
Fifth Cambridge Workshop}, Linsky, J. L. \& Stencel,
R. E. eds. Springer-Verlag (Berlin).

\bibitem[Durney, De Young \& Roxburgh(1993)]{ddr93}
{Durney}, B.~R., {De Young}, D.~S., \& {Roxburgh}, I.~W. 1993, Sol. Phys.,
  145, 207.

\bibitem[Fomalont(1981)]{fom81}
Fomalont, E.  1981, NRAO Newslett, 3, 3.

\bibitem[Geballe et al.(2001)]{gkl+01}
Geballe, T. R., et al. 2001, accepted to ApJ; astro-ph/0108443.

\bibitem[Gizis et al.(2000)]{gmr+00}
{Gizis}, J.~E., et al. 2000, AJ, 120, 1085.

\bibitem[Graham et al.(1992)]{gmg+92}
{Graham}, J.~R., et al. 1992, AJ, 104, 2016.

\bibitem[Guedel \& Benz(1993)]{gb93}
{Guedel}, M. \& {Benz}, A.~O. 1993, ApJ, 405, L63.

\bibitem[Guedel et al.(1996)]{gbs+96}
Guedel, M., et al. 1996, ApJ, 471, 1002.

\bibitem[Haisch, Strong \& Rodono(1991)]{hsr91}
{Haisch}, B., {Strong}, K.~T., \&  {Rodono}, M. 1991, ARAA, 29, 275.

\bibitem[Hawley et al.(1995)]{hfs+95}
Hawley, S.~L., et al. 1995, ApJ, 453, 464.

\bibitem[Hawley, Gizis \& Reid(1996)]{hgr96}
Hawley, S.~L., Gizis, J.~E., \& Reid, I.~N.  1996, AJ, 112, 2799.

\bibitem[Imanishi, Tsujimoto \& Koyama(2001)]{itk01}
Imanishi, K., Tsujimoto, M., \& Koyama, K. 2001, ApJ, 563, 361.

\bibitem[Johns-Krull \& Valenti(1996)]{jkv96}
Johns-Krull, C. M., \& Valenti, J. A. 1996, ApJ, 459, L95.

\bibitem[Kirkpatrick et al.(1993)]{kmh+93}
Kirkpatrick, J. D., et al. 1993, ApJ, 402, 643.

\bibitem[Kirkpatrick, Beichman \& Skrutskie(1997)]{kbs97}
{Kirkpatrick}, J.~D., {Beichman}, C.~A., \& {Skrutskie}, M.~F. 1997, ApJ,
  476, 311.

\bibitem[Kirkpatrick et al.(1999)]{kab+99}
Kirkpatrick, J. D., et al. 1999, ApJ, 519, 834.

\bibitem[Kirkpatrick et al.(2000)]{krl+00}
{Kirkpatrick}, J.~D., et al.  2000, AJ, 120, 447.

\bibitem[Krishnamurthi, Leto \& Linsky(1999)]{kal+99}
{Krishnamurthi}, A., {Leto}, G., \& {Linsky}, J.~L. 1999, AJ, 118, 1369.

\bibitem[Kundu et al.(2001)]{knw+01}
Kundu, M. R., et al. 2001, ApJ, 557, 880.

\bibitem[Leggett, Allard, \& Hauschildt(1999)]{lah98}
Legget, S. K., Allard, F., \& Hauschildt, P. H. 1998, ApJ, 509, 836.

\bibitem[Leggett et al.(2001)]{lag+01}
Leggett, S. K., et al. 2001, ApJ, 548, 908.

\bibitem[Leto et al.(2000)]{lpl+00}
{Leto}, G., et al. 2000, A\&A, 359, 1035.

\bibitem[Liebert et al.(1999)]{lkr+99}
{Liebert}, J., et al. 1999, ApJ, 519, 345.

\bibitem[Lim(1993)]{lim93}
{Lim}, J. 1993, ApJ, 405, L33.

\bibitem[Lim, White \& Slee(1996)]{lws96}
{Lim}, J., White, S.~M., \& Slee, O.~B. 1996, ApJ, 460, 976.

\bibitem[Linsky \& Gary(1983)]{lg83}
Linsky, J. L., \& Gary, D. E. 1983, ApJ, 274, 776.

\bibitem[Luhman \& Rieke(1999)]{lr99}
Luhman, K. L., \& Rieke, G. H. 1999, ApJ, 525, 440.

\bibitem[Mart{\' i}n, Rebolo \& Magazzu(1994)]{mrm94}
{Mart{\' i}n}, E.~L., {Rebolo}, R., \& {Magazzu}, A. 1994, ApJ, 436, 262.

\bibitem[Mart{\' i}n et al.(1997)]{mbd+97}
{Mart{\' i}n}, E.~L., et al. 1997, \aap, 327, L29.

\bibitem[Mart{\' i}n, Basri \& Zapatero Osorio(1999)]{mbz99}
{Mart{\' i}n}, E.~L., {Basri}, G., \& {Zapatero Osorio}, M.~R. 1999, AJ, 118,
  1005.

\bibitem[Mart{\' i}n, et al.(1999)]{mdb+99}
{Mart{\' i}n}, E.~L., et al. 1999, AJ, 118, 2466.

\bibitem[Mart{\' i}n, Brandner \& Basri(1999)]{mbb99}
{Mart{\' i}n}, E.~L., {Brandner}, W., \& {Basri}, G. 1999, Science, 283, 1718.

\bibitem[Mart{\' i}n \& Ardila(2001)]{ma01}
{Mart{\' i}n}, E.~L., \& {Ardila}, D.~R. 2001, AJ, 121, 2758.

\bibitem[Mohanty, et al.(2002)]{mbs+02}
Mohanty, S., et al. 2002, Accepted to ApJ; astro-ph/0201518.

\bibitem[Motte, Andre \& Neri(1998)]{man98}
{Motte}, F., {Andre}, P., \& {Neri}, R. 1998, \aap, 336, 150.

\bibitem[Mould et al.(1994)]{mco+94}
{Mould}, J., {Cohen}, J., {Oke}, J.~B., \& {Reid}, N. 1994, AJ, 107, 2222.

\bibitem[Neuh{\"a}user et al.(1999)]{nbc+99}
{Neuh{\"a}user}, R., it et al.  1999, A\&A, 343, 883.

\bibitem[Neupert(1968)]{neu68}
Neupert, W.~M.  1968, ApJ, 153, L59.

\bibitem[Pallavicini, Wilson \& Lang(1985)]{pwl85}
Pallavicini, R., Willson, R. F., \& Lang, K. R. 1985, A\&A, 149, 95.

\bibitem[Randich(1998)]{ran98}
Randich, S.  1998, {\it Cool Stars, Stellar Systems, and the Sun.
Tenth Cambridge Workshop}, Donahue, R. A. \& Bookbinder,
J. A. eds. ASP Conf Ser 154 (San Francisco), CD-1819.

\bibitem[Reid et al.(1999)]{rkg+99}
{Reid}, I.~N., et al. 1999, ApJ, 527, L105.

\bibitem[Reid et al.(2000)]{rkg+00}
{Reid}, I.~N., et al. 2000, AJ, 119, 369.

\bibitem[Reid et al.(2001)]{rbc+01}
{Reid}, I.~N., et al. 2001, AJ, 121, 1710. 


\bibitem[Rutledge et al.(2000)]{rbm+00}
{Rutledge}, R.~E., et al. 2000, ApJ, 538, L141.

\bibitem[Saar \& Linsky(1985)]{sl85}
{Saar}, S.~H., \& {Linsky}, J.~L. 1985, ApJ, 299, L47.


\bibitem[Schneider et al.(1991)]{sgs+91}
Schneider, D. P., et al. 1991, AJ, 102, 1180.

\bibitem[Schopper, Birk \& Lesch(2001)]{sbl01}
{Schopper}, R., {Birk}, G. T., \& {Lesch}, H. 1999, Phys. of Plasmas,
6, 4318; astro-ph/0106561. 

\bibitem[Stepanov et al.(2001)]{skz+01}
{Stepanov}, A.~V., et al. 2001, A\&A, 374, 1072.

\bibitem[Sturrock(1999)]{s99}
{Sturrock}, P.~A. 1999, ApJ, 521, 451.

\bibitem[Tinney(1993)]{tin93}
{Tinney}, C.~G. 1993, ApJ, 414, 279.

\bibitem[Tinney, et al.(1995)]{trg+95}
{Tinney}, C.~G., et al. 1995, AJ, 110, 3014.

\bibitem[Tinney, Delfosse \& Forveille(1997)]{tdf97}
{Tinney}, C.~G., {Delfosse}, X., \& {Forveille}, T. 1997, ApJ, 490, L95.

\bibitem[Tinney \& Reid(1998)]{tr98}
{Tinney}, C.~G. \& {Reid}, I.~N. 1998, MNRAS, 301, 1031.

\bibitem[Tsvetanov et al.(2000)]{tgz+00}
{Tsvetanov}, Z.~I., et al.  2000, ApJ, 531, L61.

\bibitem[Ulmschneider, Theurer \& Musielak(1996)]{utm96}
{Ulmschneider}, P., {Theurer}, J., \& {Musielak}, Z.~E. 1996, \aap, 315, 212.

\bibitem[White, Jackson \& Kundu(1989)]{wjk89}
{White}, S.~M., {Jackson}, P.~D., \& {Kundu}, M.~R. 1989, ApJS, 71, 895.

\bibitem[Wilking, Greene \& Meyer(1999)]{wgm99}
{Wilking}, B.~A., {Greene}, T.~P., \& {Meyer}, M.~R. 1999, AJ, 117, 469.

\end{thebibliography}
\end{document}